\documentclass[nofootinbib,10pt,twocolumn,preprintnumbers]{revtex4-1}
\usepackage[most]{tcolorbox}
\usepackage{slashed}

\usepackage{tikz}
\usepackage{feynmp-auto}
\usepackage{amsmath}
\usepackage[utf8]{inputenc}
\usepackage[T1]{fontenc} 
\usepackage{array}
\usepackage{bbm}
\usepackage{soul}	
\usepackage{xcolor}

\usepackage{epsfig}
\usepackage{color}
\usepackage{slashed}
\usepackage{comment}
\usepackage{epstopdf}
\usepackage{xspace}
\usepackage{mathrsfs}
\usepackage[euler]{textgreek}
\usepackage{amsmath}
\usepackage{amsthm}
\usepackage{amsfonts}
\usepackage{amssymb}
\usepackage{graphicx}
\graphicspath{ {Figures/} }
\usepackage[font={small}]{caption}
\usepackage{subcaption}
\usepackage{float}
\usepackage{multirow}
\usepackage[hang, flushmargin,bottom]{footmisc} 
\usepackage{placeins}
\usepackage{enumitem}
\usepackage{slashed}
\usepackage{bbm}
\usepackage{appendix}
\usepackage{tabularx}
\usepackage{siunitx}

\usepackage{titlesec}
\titleformat{\chapter}[display]
  {\normalfont\LARGE\bfseries}
  {\chaptertitlename\ \thechapter}{5pt}{\LARGE}
  \titlespacing*{\chapter}{0pt}{-20pt}{35pt}
\usepackage{bigstrut}
\usepackage{pst-node}
\usepackage{pstricks}
\usepackage{physics}
\usepackage{mathtools}
\usepackage{setspace}
\usepackage{fancyhdr}
\usepackage[makeroom]{cancel}
\usepackage{feyn}
\newcommand{\be}{\begin{equation}}
\newcommand{\ee}{\end{equation}}
\newcommand{\bes}{\begin{equation*}}
\newcommand{\ees}{\end{equation*}}

\usepackage{hyperref}
\hypersetup{%
  colorlinks = true,
  linkcolor  = blue,
  citecolor  = blue,
  urlcolor   = blue,
}
\usepackage{soul}
\usepackage{todonotes}
\usepackage{xpatch}
\makeatletter
\xpretocmd{\todo}{\@bsphack}{}{}
\xapptocmd{\todo}{\@esphack}{}{}
\makeatother

\newcommand{\beq}{\begin{equation}}
\newcommand{\eeq}{\end{equation}}

\definecolor{green}{HTML}{008000}
\definecolor{goldenrod}{HTML}{DAA520}
\definecolor{magenta}{HTML}{FF00FF}
\definecolor{silver}{HTML}{C0C0C0}
\definecolor{indigo}{HTML}{4B0082}
\definecolor{skyblue}{HTML}{87CEEB}
\definecolor{darkgoldenrod}{HTML}{B8860B}
\definecolor{orange}{HTML}{FFA500}
\definecolor{yellow}{HTML}{FFFF00}
\definecolor{saddlebrown}{HTML}{8B4513}
\definecolor{blue}{HTML}{0000FF}
\definecolor{turquoise}{HTML}{40E0D0}
\definecolor{yellow}{HTML}{FFFF00}
\definecolor{white}{HTML}{FFFFFF}
\definecolor{whitesmoke}{HTML}{F5F5F5}

\newcommand{\myComment}[1]{}

\usepackage{xparse}

\usetikzlibrary{decorations.pathmorphing}	
\usetikzlibrary{decorations.markings}
\tikzset{
    vector/.style={decorate, decoration={snake}, draw},
	provector/.style={decorate, decoration={snake,amplitude=2.5pt}, draw},
	antivector/.style={decorate, decoration={snake,amplitude=-2.5pt}, draw},
    fermion/.style={draw=black, postaction={decorate},
        decoration={markings,mark=at position .55 with {\arrow[draw=black]{>}}}},
    fermionr/.style={draw=black, postaction={decorate},
    decoration={markings,mark=at position .55 with {\arrow[draw=black]{<}}}},
    fermioncyan/.style={draw=black, postaction={decorate},
        decoration={markings,mark=at position .55 with {\arrow[draw=cyan]{<}}}},
    fermiondif/.style={draw=black, postaction={decorate},
        decoration={markings,mark=at position .7 with {\arrow[draw=black]{>}}}},
            fermiondif2/.style={draw=black, postaction={decorate},
        decoration={markings,mark=at position .7 with {\arrow[draw=black]{<}}}},
    fermionend/.style={draw=black, postaction={decorate},
        decoration={markings,mark=at position 1 with {\arrow[draw=black]{>}}}},
    fermionuchannel2/.style={draw=black, postaction={decorate},
        decoration={markings,mark=at position .4 with {\arrow[draw=black]{>}}}},
    scalardif/.style={dashed,draw=black, postaction={decorate},
        decoration={markings,mark=at position .7 with {\arrow[draw=black]{>}}}},
    scalarend/.style={dashed,draw=black, postaction={decorate},
        decoration={markings,mark=at position 1 with {\arrow[draw=black]{>}}}},
    fermionbar/.style={draw=black, postaction={decorate},
        decoration={markings,mark=at position .55 with {\arrow[draw=black]{<}}}},
    fermionnoarrow/.style={draw=black},
    gluon/.style={decorate, draw=black,
        decoration={coil,amplitude=4pt, segment length=5pt}},
    scalar/.style={dashed,draw=black, postaction={decorate},
        decoration={markings,mark=at position .55 with {\arrow[draw=black]{>}}}},
    scalarcyan/.style={dashed,draw=black, postaction={decorate},
        decoration={markings,mark=at position .55 with {\arrow[draw=cyan]{>}}}},
    scalaruchannel1/.style={dashed,draw=black, postaction={decorate},
        decoration={markings,mark=at position .7 with {\arrow[draw=black]{>}}}},
                  scalaruchannel2/.style={dashed,draw=black, postaction={decorate},
        decoration={markings,mark=at position .4 with {\arrow[draw=black]{>}}}},
    scalarbar/.style={dashed,draw=black, postaction={decorate},
        decoration={markings,mark=at position .55 with {\arrow[draw=black]{<}}}},
    scalarnoarrow/.style={dashed,draw=black},
    electron/.style={draw=black, postaction={decorate},
        decoration={markings,mark=at position .55 with {\arrow[draw=black]{>}}}},
	bigvector/.style={decorate, decoration={snake,amplitude=4pt}, draw},
}

\NewDocumentCommand\semiloop{O{black}mmmO{}O{above}}
{%
\draw[#1] let \p1 = ($(#3)-(#2)$) in (#3) arc (#4:({#4+180}):({0.5*veclen(\x1,\y1)})node[midway, #6] {#5};)
}

\tikzstyle{block} = [draw, rectangle, 
    minimum height=3em, minimum width=6em]

\usepackage{enumitem}
\usepackage{tikz}
\usetikzlibrary{fit}
\tikzset{%
  highlight/.style={rectangle,rounded corners,color=granate,draw,text opacity =1,
    fill opacity=0.5,thick,inner sep=0pt}
}

\usepackage{xparse}
\NewDocumentCommand\loopv{O{black}mmmO{}O{above}}
{%
\draw[#1] let \p1 = ($(#3)-(#2)$) in (#3) arc (#4:({#4+360}):({0.5*veclen(\x1,\y1)})node[midway, #6] {#5};)
}

\usepackage{placeins}

\tikzset{
    cross/.pic = {
    \draw[rotate = 45] (-#1,0) -- (#1,0);
    \draw[rotate = 45] (0,-#1) -- (0, #1);
    }
}

\tikzset{
    square/.style={%
        draw=none,
        circle,
        append after command={%
            \pgfextra \draw[#1] (\tikzlastnode.north-|\tikzlastnode.west) rectangle 
                (\tikzlastnode.south-|\tikzlastnode.east);\endpgfextra}
    },
    square/.default=black
}

\tikzstyle{block} = [draw, rectangle, 
    minimum height=3em, minimum width=6em]

\begin{document}

\title{\Large{On Gauge Theories of Neutrino Masses and Dark Matter}}
\author{Hridoy Debnath, Pavel Fileviez P\'erez}
\affiliation{
Physics Department and Center for Education and Research in Cosmology and Astrophysics (CERCA), Case Western Reserve University, Cleveland, OH 44106, USA}
\email{hxd253@case.edu,pxf112@case.edu}
%
\begin{abstract} 
We discuss the predictions in the simplest theory for neutrino masses based on the spontaneous breaking of local lepton number. This theory provides a simple theoretical framework to understand the possible relation between the origin of neutrino masses and the nature of the dark matter. In this theory, one of the fields needed for anomaly cancellation is a dark matter candidate and the local lepton number is broken at the low scale. We discuss in great detail the dark matter properties showing the allowed parameter space by the relic density bounds and the predictions for direct detection.  
The predictions for gamma and neutrino lines from dark matter annihilation are investigated. In the case of Dirac neutrinos, the bound on the effective number of relativistic degrees of freedom plays an important role and the predictions for gamma lines could be tested in the near future. We discuss the predictions in the case of Majorana neutrinos where the dark matter candidate has extra annihilation channels and compare all the predictions to the case with Dirac neutrinos.     
\end{abstract}
\maketitle
%
\section{INTRODUCTION}
%
The Standard Model (SM) describes with great precision the properties of quarks and leptons, and how they interact through the electromagnetic, weak and strong interactions. Unfortunately, the SM does not provide a mechanism to explain the origin of neutrino masses and a candidate for dark matter in the universe. 

The SM neutrinos could be Dirac or Majorana fermions. The total lepton number, $\ell=\ell_e+\ell_\mu+\ell_\tau$, is an accidental global symmetry in the SM, broken at the quantum level by $SU(2)_L$ instantons~\cite{PhysRevLett.37.8}. Here $\ell_i$ is the lepton number for each SM family. One can generate Majorana neutrino masses when the total lepton number is broken by two units. Neutrinos are Dirac fermions when $\ell$ is conserved or broken in such a way that one cannot generate any effective $\Delta \ell=\pm 2$ mass term. Majorana neutrinos are predicted in Pati-Salam~\cite{Pati:1974yy,Pati:1973rp} and grand unified theories~\cite{Fritzsch:1974nn,Georgi:1974my}. In this context, neutrino masses are generated by the seesaw mechanism~\cite{Minkowski:1977sc,Gell-Mann:1979vob,Mohapatra:1979ia,Yanagida:1979as}, typically at the $10^{14}$ GeV scale.
Unfortunately, there is no hope to test directly the origin of neutrino masses at the canonical seesaw scale.

One can consider simple gauge theories for physics beyond the SM where the total lepton number is a local gauge symmetry. Since the lepton number is not anomaly-free in the SM, one needs to add extra fermions to achieve anomaly cancelation and study the spontaneous breaking of this symmetry. Simple realistic theories based on local lepton number have been proposed in Refs.~\cite{FileviezPerez:2011pt,Duerr:2013dza,FileviezPerez:2014lnj,FileviezPerez:2024fzc}. In this context, the SM neutrinos are massive and one of the new fermions needed for anomaly cancellation is neutral and stable, then a good dark matter candidate. These theories have been discussed in detail in Refs.~\cite{FileviezPerez:2019cyn,Debnath:2023akj,Debnath:2024vpf,Schwaller:2013hqa,Madge:2018gfl,Aranda:2014zta,Carena:2022qpf} and provide a unique theoretical framework to understand the origin of neutrino masses at the low scale.

Recently, the minimal theory based on local lepton number has been proposed in Ref.~\cite{FileviezPerez:2024fzc}. In this context, the gauge anomalies are canceled with only four extra fermionic fields plus the right-handed neutrinos. This theory predicts a Majorana dark matter, one of the extra new fields is neutral and stable when it is the lightest field in the new sector. Therefore, this theory provides an unique relation between the origin of neutrino masses and the nature of dark matter. In this paper, we study in detail the properties of the dark matter candidate, showing that the local lepton number symmetry must be broken below the multi-TeV scale in order to be in agreement with the cosmological bounds on the dark matter relic density. Therefore, one can hope to test the origin of neutrino masses if this type of theory is realized in nature.

The minimal theory proposed in Ref.~\cite{FileviezPerez:2024fzc} predicts Dirac neutrinos. In this paper, we study the impact of the cosmological bounds on the effective number of relativistic degrees of freedom. We investigate in detail the constraints on the relic density and the predictions for direct detection experiments.
We show the allowed parameter consistent with all constraints and find the upper bound on the symmetry-breaking scale. We investigate the predictions for gamma and neutrino lines from dark matter annihilation. The new fermions needed for anomaly cancellation play an important role in the predictions for gamma lines. We find that the predictions for the gamma lines coming from the annihilation into a photon and the $Z$ gauge boson could be tested in the near future in gamma-ray telescopes such as CTA. 

We also study the implementation of the seesaw mechanism for Majorana neutrinos in the theory proposed in Ref.~\cite{FileviezPerez:2024fzc}. In this case, the leptophilic dark matter has several new annihilation channels and one has a different prediction for the upper bound on the symmetry-breaking scale. The theory with Majorana neutrinos predicts a new pseudo-Goldstone boson, the Majoron, that plays an important role in the prediction for dark-matter annihilation. We also discuss the predictions for direct detection and gamma lines, showing the possibility to test these predictions in the near future. This study tells us that the origin of neutrino masses and dark matter can be strongly connected in a simple theory for physics beyond the SM.

This article is organized as follows: In Sec.~\ref{minimal-theory} we discuss the minimal theory where the total lepton number is a local gauge symmetry. In Sec.~\ref{Neff} we discuss the impact of the cosmological bounds on the effective number of degrees of freedom, while in Sec.~\ref{DM} we study the properties of our dark matter candidate showing the predictions for relic density and direct detection. The predictions for gamma and neutrino lines are investigated in Sec.~\ref{Indirect}. The implementation of the seesaw mechanism for neutrino masses is discussed in Sec.~\ref{Majorana}, together with the predictions of our dark-matter candidate in this case. We summarize our main results in Sec.~\ref{Summary}.

\section{THEORY FOR LEPTON NUMBER}
\label{minimal-theory}
Recently, simple theories where the total lepton number is a local gauge symmetry~\cite{FileviezPerez:2011pt,Duerr:2013dza,FileviezPerez:2014lnj,FileviezPerez:2024fzc} have been proposed. 
These theories are based on the gauge group
\begin{equation}
SU(3)_C \otimes SU(2)_L \otimes U(1)_Y \otimes U(1)_\ell.  \end{equation}
In this article, we will discuss the predictions for the dark matter candidate present in the minimal theory where all gauge anomalies are canceled with only four fermionic representations. In this context, the gauge anomalies are canceled with the fermionic fields~\cite{FileviezPerez:2024fzc}:
\begin{eqnarray}
\Psi_L &\sim& ({\bf{1}},{\bf{1}},-1,3/4), \    \Psi_R \sim ({\bf{1}},{\bf{1}},-1,-3/4), \\
\chi_L &\sim& ({\bf{1}},{\bf{1}},0,3/4), \ 
{\rm{and}} \ \rho_L \sim ({\bf{1}},{\bf{3}},0,-3/4).
\end{eqnarray}
plus the right-handed neutrinos, 
$
\nu_R^i \sim ({\bf{1}},{\bf{1}},0,1),
$
with $i=1,2,3$.
We can generate masses for the new fields using the new Yukawa interactions:
\begin{eqnarray}
- \mathcal{L} &\supset& \lambda_\rho {\rm{Tr}} (\rho_L^T C \rho_L) S + \lambda_\Psi \bar{\Psi}_L \Psi_R S + \lambda_\chi \chi_L^T C \chi_L S^* \nonumber \\
&+& \rm{h.c.},   
\end{eqnarray}
where $S\sim ({\bf{1}},{\bf{1}},0,3/2)$. 
The SM leptons have the following interactions:
\begin{eqnarray}
- \mathcal{L} \supset Y_\nu  \bar{\ell}_L i \sigma_2 H^* \nu_R \ + \
y_e  \bar{\ell}_L H e_R
+\rm{h.c.},
\end{eqnarray}
Notice that only one extra scalar field, $S$, with lepton number $3/2$ is needed to generate masses for the new fermions and the neutrinos are predicted to be Dirac fermions.  
The scalar fields can be written as 
\begin{eqnarray}
H&=&\begin{pmatrix}
h^+\\
\frac{1}{\sqrt{2}}(v_{0} + h_0) e^{i \sigma_0/v_0}
\end{pmatrix}, 
\end{eqnarray}
and
\begin{eqnarray}
S &=& \frac{1}{\sqrt{2}}\left(v_{S} + h_S \right) e^{i \sigma_S/v_S}.
\label{S-filed}  
\end{eqnarray}
After symmetry breaking, $v_0\neq0$ and $v_S\neq0$, the physical neutral CP-even Higgses are defined as
\begin{eqnarray}
h &=& h_0 \cos \theta_\ell - h_S \sin \theta_\ell,\\
h_\ell &=& h_0 \sin \theta_\ell  + h_S \cos \theta_\ell.
\end{eqnarray}
 This theory predicts a new neutral gauge boson that couples mainly to the SM leptons and the new fermions needed for anomaly cancellation. The new gauge boson acquires mass after the $U(1)_\ell$ gauge symmetry is broken and its mass is given by
\begin{equation}
    M_{Z_\ell}=\frac{3}{2} g_\ell v_S.
\end{equation}
After the spontaneous breaking of lepton number one can see that this theory predicts two extra electrically charged physical fermions: 
\begin{equation}
\Psi^-=\Psi^-_L + \Psi^-_R \ 
{\textrm{and}} \ 
\rho^-=\rho_L^-+(\rho_L^+)^C, \nonumber
\end{equation}
and two neutral Majorana fields: 
\begin{equation}
\chi=\chi_L + (\chi_L)^C \ {\textrm{and}} \ \rho^0=\rho_L^0+(\rho_L^0)^C.
\nonumber
\end{equation}
The new fermions masses are given by 
\begin{eqnarray}
M_{\rho^0} &=& \sqrt{2} \lambda_\rho v_S, \
M_{\rho^-} = M_{\rho^0} + \delta M, \\
M_\chi &=& \sqrt{2} \lambda_\chi v_S, \ {\rm{and}} \
M_\Psi = \lambda_\Psi \frac{v_S}{\sqrt{2}}.
\end{eqnarray}
The new physical charged fermion $\rho^-$ can decay into the neutral fields, $\rho^0$, and a charged pion $\pi^-$, due to the fact that a very small splitting is generated at one-loop level ~\cite{Cirelli_2006}.

The lightest new field in the new sector is stable because after the spontaneous breaking of $U(1)_\ell$, the theory has an accidental discrete symmetry
\begin{eqnarray}
{\mathcal{Z}_2}: \Psi_L \to - \Psi_L, \Psi_R \to - \Psi_R, \rho_L \to - \rho_L, \chi_L \to - \chi_L. \nonumber     
\end{eqnarray}
 protecting the stability of the lightest new fermion.
In this case the lightest field between $\rho^0$ and $\chi$, can be a dark matter candidate. Since the stability of the dark matter candidate is very important we must think about possible higher-dimensional operators that could modify the predictions at the renormalizable level. The local gauge symmetry allows the following higher-dimensional operators
\begin{equation}
- \mathcal{L} \supset 
\frac{y}{\Lambda} H^\dagger \chi_L^T C \rho_L H +
\frac{c_e}{\Lambda^2} \bar{\Psi}_L e_R \bar{\nu}_R \chi_L + {\rm{h.c.}}.
\end{equation}
The first dimension five operator tells us that we have only one stable neutral field in the theory, while the dimension six operator tells us that the new charged field $\Psi$ decays into the SM leptons and $\chi$. 
It is important to emphasize that this minimal theory for spontaneous breaking of local lepton number makes two main predictions: a) the SM neutrinos are Dirac fermions, and b) one has a Majorana dark matter candidate. 
%
\section{COSMOLOGICAL BOUNDS: $N_{eff}$}
\label{Neff}
In the theory discussed above one predicts the presence of right-handed neutrinos. In this context, the SM neutrinos are Dirac fermions and one needs to understand the implications of the cosmological bounds on $N_{eff}$.
The SM predicts that $N_{eff}^{SM}=3.043$~\cite{Cielo:2023bqp}, 
while from $\Lambda CDM$ one has $N_{eff}=2.99\pm0.17$~\cite{Planck:2018vyg}.
In this theory, one has extra relativistic degrees of freedom, three copies of right-handed neutrinos.
The right-handed neutrinos will decouple from the thermal bath when the interaction rate drops below the expansion rate of the universe. If the right-handed neutrinos decouple at temperature $T_{\nu_R}^{dec}$, then at $T_{\nu_R}^{dec}$,
\begin{eqnarray}
    \Gamma (T_{\nu_R}^{dec}) &=& H (T_{\nu_R}^{dec}) \nonumber \\
    &=& \sqrt{\frac{4 \pi^3 G_N}{45} \left(g(T_{\nu_R}^{dec})+\frac{21}{4} \right)} (T_{\nu_R}^{dec})^2.
\end{eqnarray}
Here $g(T)$ is the number of SM  relativistic degrees of freedom at temperature T, $G_N =  1/M_{pl}^2$  and the factor $21/4$ reflects the degrees of freedom for three relativistic right-handed neutrinos. The right-handed neutrino pair can annihilate into the SM degrees of freedom via exchange of $Z_{\ell}$ and remain in thermal equilibrium. Then the interaction rate  per $\nu_R$ can be written as 
\begin{equation}
    \Gamma_{\nu_R} = \sum_i n_{\nu_R} \left<\sigma v (\nu_R \bar{\nu}_R \rightarrow f_i \bar{f}_i)\right>.
\end{equation}
Here $g_{\nu_R} =2 $ for right-handed neutrinos and $n_{\nu R}$ is the number density of a single flavor of right-handed neutrinos and antineutrinos, which can be written as
\begin{figure}[h]
         \centering         \includegraphics[width=0.45\textwidth]{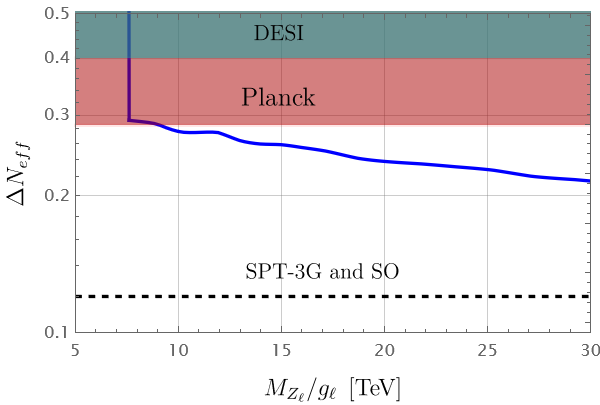}
        \caption{ 
        Extra contribution to the number of relativistic degrees of freedom as a function of $M_{Z_\ell}/g_\ell$. Here the red and cyan shaded regions are the bound from Planck~\cite{Planck:2018vyg} and DESI~\cite{desicollaboration2024desi2024vicosmological} experiments, respectively. The projected SPT and SO bounds~\cite{Ade_2019,Benson_2014} are shown by the dashed line.}
        \label{NeffBound}
\end{figure}
\begin{equation}
    n_{\nu_R } (T) = g_{\nu_R} \int \frac{d^3 p}{(2 \pi)^3 }f_{\nu_R}(p) = \frac{3}{2 \pi^2} \zeta(3) T^3.
\end{equation}
Here, $f_{\nu_R}(p)$ is the Fermi-Dirac distribution and $\zeta(x)$ is the Riemann zeta function. If the right-handed neutrinos collide with four-momentum $p^{\mu}$ and $k^{\mu}$ with a relative angle $\theta$, then the interaction rate per neutrino is written as 
\begin{eqnarray}
   \Gamma(T)  &=& \frac{g_{\nu_R}^2}{n_{\nu_R}(T)} \int \frac{d^3 p}{(2 \pi)^3} \frac{d^3 k}{(2 \pi)^3}f_{\nu_R}(p) f_{\nu_R}(k) \sigma_f(s)v_M.
   \nonumber \\
\end{eqnarray}
Here, $v_M = 1- \cos{\theta}$ is the Moller velocity, and $s = 2 pk(1-\cos{\theta})$ is the center-of-mass energy. The annihilation of right-handed neutrinos into two SM fermions is given in Eq.(\ref{nuRnuRff}). The new gauge boson is heavy, $M_{Z_{\ell}}\gg T_{\nu_R}^{dec}$, and one can consider the limit $s \ll M_{Z_\ell}$. Neglecting fermion masses, one finds
\begin{eqnarray}
    \Gamma_{\nu_R} (T) = \frac{49 \pi^5 T^5}{97200 \zeta(3)} \left( \frac{g_\ell}{M_{Z_\ell}}\right)^4 \sum_f N_f^C n_f^2.
\end{eqnarray}
Here, $N_f^C$ is the charge of the fermions and $n_f$ is the charge of the fermions under $U(1)_{\ell}$ symmetry.
Now, we can write 
\begin{eqnarray}
    (T^{dec}_{\nu_R})^3 &=& \sqrt{\frac{4 \pi^3 G_N}{45} \left(g(T)+\frac{21}{4} \right)} \hspace{0.2 cm}\frac{97200 \zeta(3)}{49 \pi^5} \hspace{0.2 cm} \nonumber \\
    & \times &
    \frac{1}{ \sum_f N_f^C n_f^2 } \left(\frac{M_{Z_\ell}}{g_\ell}\right)^4.
\end{eqnarray}
Knowing the decoupling temperature for the right-handed neutrinos for a given value of the new gauge boson mass and coupling, we can study the constraints from the effective degree of relativistic degrees of freedom.
The extra contribution to $N_{eff}$, $\Delta N_{eff}$, can be written as
\begin{eqnarray}
    \Delta N _{eff} =  N_{\nu_R} \left( \frac{ g(T^{dec}_{\nu_L})}
    {g(T^{dec}_{\nu_R)}}\right)^{4/3}. 
\end{eqnarray}
Here $N_{\nu_R} =3 $, \hspace{0.1 cm} $g(T^{dec}_{\nu L}) \approx 43/4$ and $g(T^{dec}_{\nu R})$ is the number of SM relativistic degrees of freedom at temperature $T^{dec}_{\nu_R}$.
In Fig.\ref{NeffBound} we show the predictions for $\Delta N_{eff}$ as a function of the ratio between the leptophilic gauge boson mass and its gauge coupling. As one can appreciate, the $N_{eff}$ bound is much stronger than the LEP bound, $M_{Z_\ell}/g_\ell \gtrsim 7$ TeV, and excludes the region below $ M_{Z_{\ell}}/g_{\ell} < 9$ TeV. The sharp drop in the number of relativistic degrees of freedom at $M_{Z_\ell}/g_\ell\approx  8$ TeV is due to the QCD phase transition and we followed the calculation from Ref.~\cite{Husdal_2016}.
These bounds are very important to understand the allowed parameter space when we study the predictions for the dark matter relic density in the next section. The dashed line in Fig.~\ref{NeffBound} shows the projected SPT and SO experimental bounds~\cite{Ade_2019,Benson_2014}. Notice that in the future one could test or rule out these predictions even if the ratio, $M_{Z_\ell}/g_\ell$, is above 30 TeV, far from the reach of any proposed collider experiment.
%
\section{LEPTOPHILIC DARK MATTER}
\label{DM}
 The lightest field between $\chi^0$ and $\rho^0$ is stable and can be a dark matter candidate. We focus our study in the case where the dark matter is the Majorana, $\chi=\chi_L+(\chi_L)^C$, because it is a generic candidate that appears in the different theories proposed where the total lepton number is a local gauge symmetry.
\begin{itemize}
\item {\textit{Relic density constraints:}}
\label{SecRelic}
In this theory, the dark matter candidate has the following annihilation channels:
\begin{equation}
 \chi \chi \to \bar{\nu}_i \nu_i, e^+_i e^-_i, Z_\ell Z_\ell, h_i Z_\ell, h_i h_j, WW,ZZ,..   
\end{equation}
See appendix~\ref{Annihilation-Dirac} for the Feynman graphs for the dark matter annihilation channels.
\begin{figure} [h]  
        \centering              \includegraphics[width=0.52\textwidth]{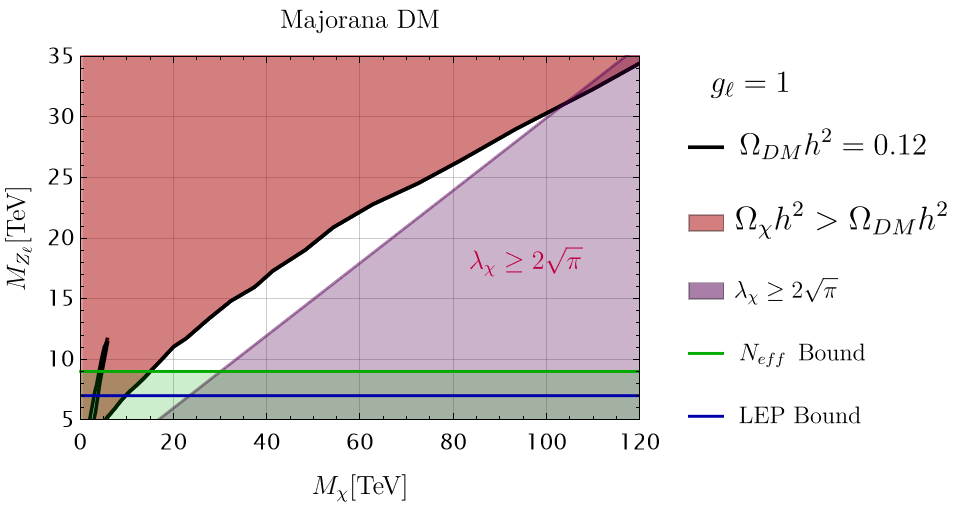}
         \caption{Allowed parameter space by the dark matter relic density constraints and perturbative bound in the $M_{Z_\ell}$ - $M_{\chi}$ plane when $g_{\ell}=1.0$ and $M_{h_\ell} = 200$ GeV. Here we consider the Sommerfeld enhancement due to the interaction between $\chi$ and $h_\ell$. The black line corresponds to the scenarios with relic density $\Omega_\chi h^2 = 0.12$. The red-shaded region is excluded by the overproduction of dark matter and the purple shaded region is excluded by the perturbative bound on $\lambda_\chi$. The green and blue lines show the bounds from $\Delta N_{eff}$ and LEP, respectively. }
         \label{Relic1}
\end{figure}
Notice that, in this theory, one needs only a few free parameters to study the dark matter relic constraint. The relevant free parameters are 
$$M_{\chi},\hspace{0.2 cm}M_{Z_\ell},\hspace{0.2 cm}M_{h_\ell},\hspace{0.2 cm}g_{\ell},\hspace{0.2 cm} \textrm{and} \ \sin{\theta_\ell}.$$
 
In Fig.~\ref{Relic1} using the standard freeze-out calculation, we show the allowed parameter space in $M_{Z_\ell}-M_\chi$ plane when $g_\ell=1.0$. The parameter space has two regions a)  the resonance region when $M_{Z_\ell} \approx 2 M_\chi $ and b) the non-resonant region (most generic). The small region below $M_\chi \approx 10$ TeV shows the resonant region where the main contribution comes from the $\chi \chi \to e_i^+ e_i^- , \nu_i \bar{\nu_i}$ annihilation channels. Notice that those channels are velocity suppressed and the resonant region is almost excluded by the $N_{eff}$ bound. In the non-resonant region, the main contribution comes from the $\chi \chi \to Z_\ell Z_\ell$ and $\chi \chi \to \ Z_\ell h_\ell $ annihilation channels. This region tells us the maximum allowed value by the new gauge boson that satisfies the cosmological relic bounds. In this case, the maximum value for $M_{Z_\ell}$ is around 33 TeV, and above that the region is excluded by the perturbative bound on the Yukawa coupling, $\lambda_\chi$. The $\chi \chi \to h h_\ell,\hspace{0.1 cm}  hh, \hspace{0.1 cm} W W , \hspace{0.1 cm} ZZ$ channels are suppressed by the mixing angle $\sin{\theta_\ell}$, while the $\chi \chi \to h_\ell h_\ell$ channel is suppressed by the dark-matter velocity. One can appreciate the fact that in this theory the symmetry-breaking scale, or $M_{Z_\ell}$, must be below the multi-TeV scale in agreement with all the experimental bounds. 
\begin{figure} [b]  
        \centering       
        \includegraphics[width=0.45\textwidth]{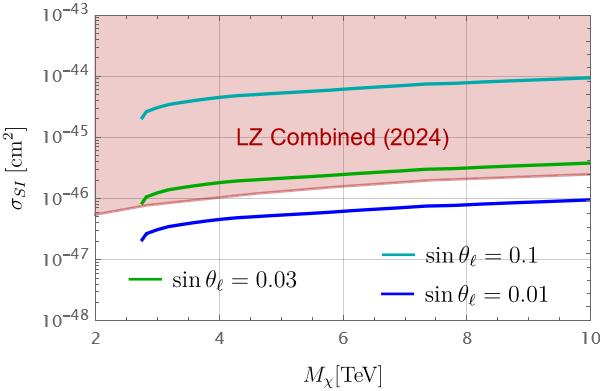}
         \caption{Spin-independent dark matter-nucleon cross-section as a function of dark matter mass when  $M_{h_\ell}=200$ GeV. The red-shaded region is excluded by the LZ experiment ~\cite{aalbers2024darkmattersearchresults}. Here we used $f_N=0.308$~\cite{Hoferichter_2017}.}
         \label{DD}
\end{figure}
%
\item{\textit{Direct Detection:}}
The dark matter-nucleon interaction is mediated by SM Higgs and the new Higgs $h_{\ell}$. The spin-independent cross section can be written as 
\begin{eqnarray}
\sigma_{\chi N}^\text{SI}(h_i)&=&\frac{9 G_F}{2\sqrt{2}  \pi}\sin^2 \theta_{\ell} \cos^2 \theta_{\ell} M_N^4 \nonumber \\
&\times& \frac{g_{\ell}^2 M_\chi^2}{M_{Z_\ell}^2}\left(\frac{1}{M_{h}^2}-\frac{1}{M_{h_\ell}^2}\right)^2 f_N^2.
\end{eqnarray}
Here $G_F$ represents the Fermi constant and $M_N$ is the nucleon mass. In Fig.~\ref{DD} we show the prediction of the spin independent dark matter-nucleon cross section as a function of dark matter mass when  the mixing angle between the Higgses, $\sin{\theta}$ is $0.1$, $0.03$ and $0.01$. Here all the points satisfy the cosmological relic density constraint, $\Omega_{DM} h^2=0.12$. The red shaded region is excluded by the new LUX-ZEPLIN (LZ)~\cite{aalbers2024darkmattersearchresults} experimental bounds. Therefore, one can say that if the mixing angle between the SM Higgs and $h_\ell$ is smaller than $0.03$ one can easily satisfy the direct detection bounds. Unfortunately, all these scenarios are ruled out by the $N_{eff}$ bounds discussed in the previous section. It is interesting to emphasize that in this case the $N_{eff}$ bounds are much stronger than the direct detection bounds when neutrinos are Dirac fermions. 

The leptophilic gauge boson $Z_\ell$ can also mediate the electron-dark matter interaction but the collider bounds on the $Z_\ell$ mass are very strong, $M_{Z_\ell}/g_\ell \gtrsim 7$ TeV, and one can easily satisfy the experimental bounds on the electron-dark matter interactions. One can also generate at one-loop level the $Z_\ell$ couplings to the SM quarks, since the $Z_\ell$ is heavy these couplings are highly suppressed.
\end{itemize}
%
\section{INDIRECT DETECTION}
\label{Indirect}
In this theory one can have interesting signatures from the dark matter annihilation. Here we discuss the predictions for gamma and neutrinos lines.
\begin{itemize}
\item{Gamma Lines}: 
\begin{itemize}
\item $\chi \chi \to \gamma \gamma$:
In this theory, dark matter can annihilate into leptons,  massive gauge bosons, and Higgses at tree level. However, at one loop level one can have the dark matter annihilation into two photons or into a photon and the $Z$ gauge boson. As it is well-known, gamma rays can play a significant role in the detection of dark matter as they can travel almost unperturbed throughout the galaxy and therefore, one can trace back to their origin. 

In this theory, one can generate an effective interaction between $Z_\ell$ and two photons, where inside the loop one has the new fields needed for anomaly cancellation and the SM charged leptons. However, the new fermions, $\Psi^-$ and $\rho^-$, couple to the $Z_\ell$ with a axial vector coupling and they can give rise to a non-velocity suppressed cross section for the annihilation into two photons. For a generic study of the gamma lines in dark matter models with new gauge symmetries see the study in Ref.~\cite{FileviezPerez:2019rcj}.
See Fig.~\ref{FGlines} for the relevant Feynman diagrams. The energy of the gamma line is $E_\gamma = M_\chi$ and the annihilation cross section is given in Eq.(\ref{Gamma1}).
\begin{widetext}
\begin{figure}[h]
\centering
\begin{eqnarray*}    
\begin{gathered}
\begin{tikzpicture}[line width=1.5 pt,node distance=1 cm and 1.5 cm]
\coordinate[label =left: $\chi$] (i1);
\coordinate[below right= 1cm of i1](p1);
\coordinate[below left = 1cm of p1, label= left:$\chi$](i2);
\coordinate[right=0.5 cm of p1](j1);
\coordinate[right=0.6 cm of j1, label=$Z_\ell$](vaux);
\coordinate[right = of j1](p2);
\coordinate[right=0.75 cm of p2](vmare);
\coordinate[above=0.505 cm of vmare ,label={[rotate=32]{$\psi^-,\, \rho^-$}}](vaux1);
\coordinate[below=1.1cm of vmare](vaux2);
\coordinate[right= 1cm of vmare](vaux3);
\coordinate[above right = of p2](v1a);
\coordinate[right = of v1a, label=right: $\gamma$] (v1);
\coordinate[below right = of p2](v2a);
\coordinate[right = of v2a,label=right: $\gamma \text{, }Z$] (v2);
\draw[fermionnoarrow] (i1) -- (j1);
\draw[fermionnoarrow] (i2) -- (j1);
\draw[vector] (j1) -- (p2);
\draw[vector] (v1a) -- (v1);
\draw[fermion] (v1a)--(v2a);
\draw[fermion](p2)--(v1a);
\draw[fermion] (v2a)--(p2);
\draw[vector] (v2a) -- (v2);
\draw[fill=cyan] (j1) circle (.1cm);
\draw[fill=cyan] (p2) circle (.1cm);
\draw[fill=red] (v1a) circle (.1cm);
\draw[fill=red] (v2a) circle (.1cm);
\end{tikzpicture}
\end{gathered}
\begin{gathered}
\begin{tikzpicture}[line width=1.5 pt,node distance=1 cm and 1.5 cm]
\coordinate[label =left: $\chi$] (i1);
\coordinate[below right= 1cm of i1](p1);
\coordinate[below left = 1cm of p1, label= left:$\chi$](i2);
\coordinate[right=0.5 cm of p1](j1);
\coordinate[right=0.6 cm of j1, label=$h_i$](vaux);
\coordinate[right = of j1](p2);
\coordinate[right=0.75 cm of p2](vmare);
\coordinate[above=0.5 cm of vmare,,label={[rotate=32]{$\psi^-,\, \rho^-$}}](vaux1);
\coordinate[below=1.1cm of vmare](vaux2);
\coordinate[right= 1cm of vmare](vaux3);
\coordinate[above right = of p2](v1a);
\coordinate[right = of v1a, label=right: $\gamma$] (v1);
\coordinate[below right = of p2](v2a);
\coordinate[right = of v2a,label=right: $\gamma \text{, }Z$] (v2);
\draw[fermionnoarrow] (i1) -- (j1);
\draw[fermionnoarrow] (i2) -- (j1);
\draw[scalarnoarrow] (j1) -- (p2);
\draw[vector] (v1a) -- (v1);
\draw[fermion] (v1a)--(v2a);
\draw[fermion](p2)--(v1a);
\draw[fermion] (v2a)--(p2);
\draw[vector] (v2a) -- (v2);
\draw[fill=cyan] (j1) circle (.1cm);
\draw[fill=cyan] (p2) circle (.1cm);
\draw[fill=red] (v1a) circle (.1cm);
\draw[fill=red] (v2a) circle (.1cm);
\end{tikzpicture}
\end{gathered}
\end{eqnarray*}
\caption{Feynman graphs for the gamma lines.}
\label{FGlines}
\end{figure}
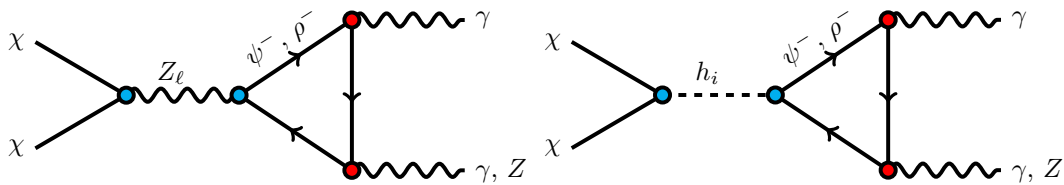
\end{widetext}

The interaction between the new fermions needed for anomaly cancellation and the leptophilic gauge boson can be written as
\begin{equation}
{\cal L} \supset g_\ell n_A^{F} \ \overline{F} \gamma^\mu \gamma^5 F Z^\ell_\mu, 
\end{equation}
where $F=\psi^- \ \text{or} \ \rho^-$, while 
$n_A^{\Psi^-}= 3/4$ and $n_A^{\rho^-}= -3/4$. 
\begin{figure} [h]  
        \centering              \includegraphics[width=0.45\textwidth]{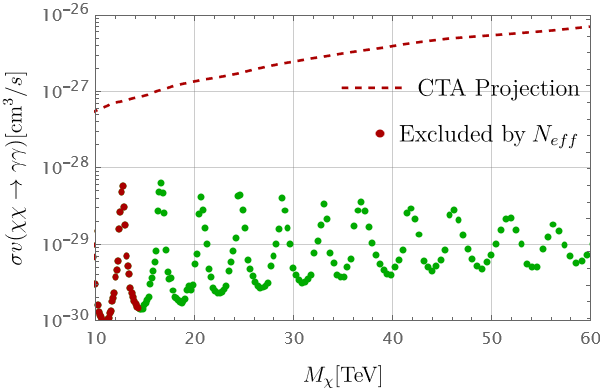}
         \caption{Thermal average cross-section for $\chi \chi \to \gamma \gamma $ as a function of dark matter mass, $M_{\chi}$. Here all the points satisfy the relic density constraint, $\Omega_{DM} h^2=0.12$, and the red points are excluded by $N_{eff}$ bound. For the numerical results we use $M_{h_\ell}=200$ GeV, $g_\ell=1.0$ , $M_{\Psi^-}=1.2 M_\chi$ and $M_{\rho^-}=3 M_\chi$. The dashed line shows the projected CTA bounds~\cite{CTAO:2024wvb}. }
         \label{chichigammagamma}
\end{figure}

The predictions for $\sigma v (\chi \chi \to \gamma \gamma)$ as a function of $M_\chi$ are shown in Fig.~\ref{chichigammagamma}. Here the red points are excluded by the $N_{eff}$ bounds discussed above. Notice that the $N_{eff}$ bounds exclude scenarios with $M_\chi \leq 14 $ TeV. The red-shaded region shows the projected CTA bounds~\cite{CTAO:2024wvb}. It is interesting to note that the projected CTA bounds can be close to the predictions, but still one needs to improve the experimental bounds by one order of magnitude. It is important to mention that the Sommerfeld enhancement due to the $h_\ell\chi\chi$ interaction is very important to predict the results shown in Fig.~\ref{chichigammagamma}.
See Appendix~\ref{SecSom} for the discussion about the Sommerfeld enhancement in this theory.
\item{$\chi \chi \to Z \gamma$:}
\label{secZgamma}
\begin{figure} [h]  
        \centering       
        \includegraphics[width=0.45\textwidth]{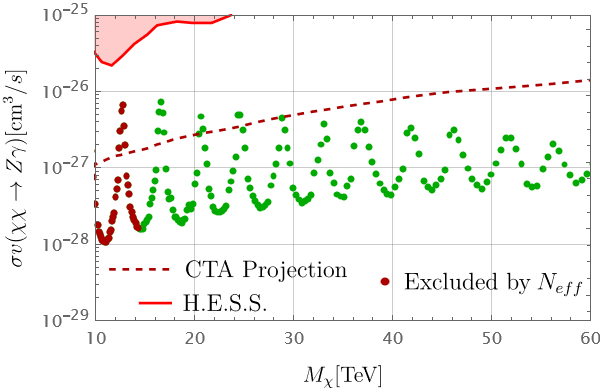}
         \caption{Thermal average cross section for $\chi \chi \to Z \gamma $ as a function of dark matter mass, $M_{\chi}$. Here all scenarios satisfy the relic density constraint, $\Omega_{DM} h^2=0.12$, and the red points are excluded by the $N_{eff}$ bound. Here we use as input parameters $M_{h_\ell}=200$ GeV, $ g_\ell=1.0$, $M_{\Psi^-}=1.2 M_\chi$ and $M_{\rho^-}=3 M_\chi$. The red shaded region is excluded by the H.E.S.S. ~\cite{Abramowski_2013} experiment and the dashed line shows the projected CTA experimental bounds~\cite{CTAO:2024wvb}. }
         \label{INDZgamma}
\end{figure}
The dark matter can also annihilate into $Z \gamma$ at one loop label as shown in Fig.~\ref{FGlines}.  In this case, the energy of the gamma line is given by
\begin{equation}
E_\gamma = M_\chi \left(1 - \frac{M_Z^2}{4 M_{\chi}^2} \right),
\label{Eq21}
\end{equation} 
and the annihilation cross section is given in Eq.(\ref{chichiZgamma}). The interactions between the new charged fermions and the Z gauge boson can be written as
\begin{equation}
{\cal L} \supset  - \frac{e}{\sin \theta_W \cos \theta_W} g_V^F  \hspace{0.1 cm}\overline{F} \gamma^\mu  F Z_\mu,
\end{equation}
with $g_V^{\Psi^-}=\sin^2 \theta_W$ and $g_V^{\rho^-}=- \cos^2 \theta_W$.

In Fig.~\ref{INDZgamma}, we show the predictions of $\sigma v (\chi \chi \to Z \gamma)$ as a function of dark matter mass $M_\chi$. In the case of $\sigma (\chi \chi \to \gamma \gamma)$, there is a cancelation between the contributions of the two extra fields, $\rho^-$ and $\Psi^-$, because they have oppositive couplings to the gauge boson $Z_\ell$. In the case of the annihilation to $Z\gamma$ this cancellation is absent because the couplings of $\rho^-$ and $\Psi^-$ to the $Z$ gauge boson are quite different. Therefore, the cross section for $\chi \chi \to Z \gamma$ can be much larger. Notice that the projected CTA experimental bounds~\cite{CTAO:2024wvb} can be very close to the numerical predictions and one can hope to test this theory for dark matter in the near future. The scenarios shown in red are excluded by the $N_{eff}$ bounds but most of the scenarios are allowed and one could test the predictions at CTA~\cite{CTAO:2024wvb}.

\end{itemize}

\item{Contributions to the continuum:}
\begin{itemize}
\item{$\chi \chi \to Z_{\ell}h_{\ell} \to e_i^+e_i^- \gamma \gamma:$}
\label{Boxgamma}
This theory also predicts gamma-ray spectrum through cascade processes, commonly known as box-shaped gamma-ray spectrum. As mentioned earlier, in the non-resonant region, the annihilation channel $\chi \chi \to Z_\ell h_\ell$ dominates and $h_\ell $ can decay into two photons, and this channel could spoil the visibility of the gamma lines discussed above. Although the annihilation channel $\chi \chi \to Z_\ell h_\ell$ dominates in the non-resonant region, due to a small branching ratio for the decay $h_\ell \to \gamma \gamma$, the cross-section of $\chi \chi \to Z_\ell h_\ell \to e_i^+e_i^- \gamma\gamma$ is smaller than the cross-section for the gamma lines. 

\item{$\chi \chi \to 4 \gamma$}: One can also have the box-shaped gamma-ray spectrum process, $\chi \chi \to h_\ell h_\ell \to 4 \gamma$. However, the branching ratios of $h_{\ell} \to \gamma \gamma$ is very small (order of $10^{-5}$, and the contribution of these channels can be neglected. See Ref.~\cite{Debnath:2024cil} for a detailed discussion of these processes.

\item{$\chi \chi \to e_i^+ e_i^- \gamma:$}
In this theory, one can have final state radiation processes predicting a photon continuum spectrum from dark matter annihilation that could destroy the visibility of the gamma lines. It is important to note that the channels mediated by the SM Higgs and the new Higgs are suppressed by the small Yukuwa couplings and are also velocity suppressed. 
Thus the leading contribution for the continuum spectrum comes from the channels mediated by the new gauge boson $Z_\ell$. The amplitude square for the channels mediated by $Z_\ell$ can be written as 
\begin{equation}
|{\cal M}|^2_\text{FSR} = \frac{M_\ell^2}{M^2_{Z_\ell}} A +  v^2 B + {\cal O}(v^4),
\label{FSR1}
\end{equation}
where the coefficients $A$ and $B$ are given in Eqs.(\ref{A}) and (\ref{B}).
Notice that the leading term for the final state radiation is highly suppressed by the ratio $(M_\ell/M_{Z_\ell})^2$ where $M_\ell$ is the mass of the charged lepton. In the next subsection, we will 
show the correlation between the predictions for gamma lines and the processes contributing to the photon continuum spectrum. 
\end{itemize}

The photon flux  for $\chi \chi \to \gamma \gamma$ is given by
\begin{eqnarray}
\frac{d\Phi_{\gamma \gamma}}{dE_\gamma}&=& \frac{n_\gamma }{8 \pi M_\chi^2} \frac{ d ( \sigma v_\text{rel} (\chi \chi \to \gamma \gamma))}{dE_\gamma}J_\text{ann} \nonumber \\
&=& \frac{n_\gamma ( \sigma v_\text{rel} (\chi \chi \to \gamma \gamma) )}{8\pi M_\chi^2}\frac{dN_{\gamma \gamma}}{dE_\gamma} J_\text{ann},
\end{eqnarray}
Here $n_\gamma=2$, $J_\text{ann}$ is the astrophysical factor that takes into account the dark matter distribution in the galaxy and the spectrum function can be written as
\begin{equation}
\frac{dN_{\gamma \gamma}}{dE_\gamma}=\int_0^\infty dE_0 \,  \delta(E_0 - M_\chi) \, G(E_\gamma, \xi /\omega, E_0),
\end{equation}
Here, a Gaussian function is used to model the detector resolution, $G(E_\gamma,\xi/\omega, E_0)$, which reads as
\begin{equation}
G(E_\gamma,\xi/\omega, E_0)=\frac{1}{\sqrt{2\pi}E_0(\xi/\omega)} e^{ -\frac{(E_\gamma - E_0)^2}{2E_0^2(\xi/\omega)^2}},
\end{equation}
where $\omega = 2 \sqrt{2 \rm{log} 2} \approx 2.35$ determines the full width at half maximum, with the standard deviation given by $\sigma_0=E_0 \xi/w$. and  $\xi$ represents the detector energy resolution.  
The flux for the gamma lines from the annihilation channel $\chi \chi \to Z \gamma$ can be written as
\begin{eqnarray}
\frac{d\Phi_{\gamma Z}}{dE_\gamma}&=& \frac{n_\gamma }{8 \pi M_\chi^2} \frac{ d ( \sigma v_\text{rel} (\chi \chi \to \gamma Z))}{dE_\gamma}J_\text{ann} \nonumber \\
&=&\frac{n_\gamma ( \sigma v_\text{rel} (\chi \chi \to \gamma Z))}{8\pi M_\chi^2}\frac{dN_{\gamma Z}}{dE_\gamma} J_\text{ann},
\end{eqnarray}
Here $n_\gamma=1$ and
\begin{equation}
\frac{dN_{\gamma Z}}{dE_\gamma}=\int_0^\infty dE_0 \,  W_{\gamma Z} \, G(E_\gamma, \xi /\omega, E_0),
\label{spectral}
\end{equation}
where the spectral function $W_{\gamma Z}$ is written as 
\begin{equation}
W_{\gamma Z} = \frac{1}{\pi}\frac{4M_\chi M_Z \Gamma_Z}{(4M_\chi^2-4M_\chi E_0 - M_Z^2)^2+ \Gamma_{Z}^2M_Z^2}.
\label{BWS}
\end{equation}
\begin{figure} [h]  
        \centering       
        \includegraphics[width=0.55\textwidth]{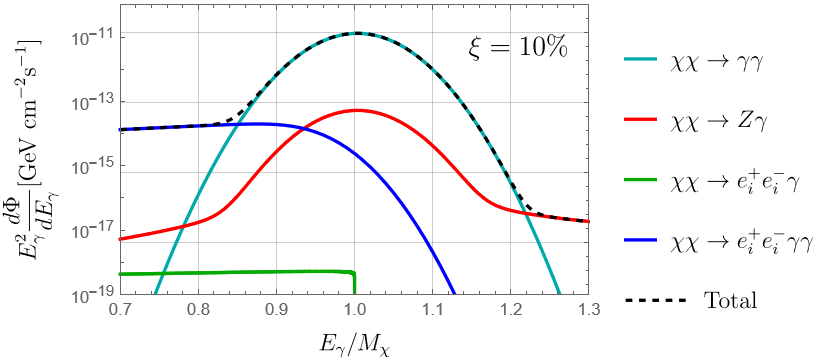}
         \caption{Total photon flux as a function of $E_\gamma/M_{\chi}$ with $10\%$ energy resolution. Here we use $M_{h_\ell}=200$ GeV, $g_{\ell}=1.0$, $M_{\chi}=36$ TeV, $M_{Z_\ell}=16.5$ TeV, $M_{\rho^-}=3 M_\chi$ and $M_{\Psi^-}=1.2 M_\chi$.}
         \label{EnSpec}
\end{figure}
Here $M_Z$ and $\Gamma_Z$ are the mass and the decay width of the $Z$ gauge boson, respectively.
Notice that for the benchmark scenario shown in Fig. ~\ref{EnSpec}, although the cross section for $\chi \chi \to Z \gamma$ is larger than the $\chi \chi \to \gamma \gamma $, the photon flux is lower due to the suppression of the spectral function, $W_{\gamma Z}$. 

The flux for the box-shaped  gamma-ray annihilation channel $\chi \chi \to e_i^+e_i^-\gamma \gamma$ is given by
\begin{equation}
\frac{d\Phi_{ \gamma }}{dE_\gamma}
=\frac{ 2 ( \sigma v_\text{rel} (\chi \chi \to Z_\ell h_\ell)) BR (h_\ell \to 2 \gamma)}{8\pi M_\chi^2}\frac{dN_{2\gamma}}{dE_\gamma} J_\text{ann},
\end{equation}
where
\begin{equation}
\frac{dN_{2 \gamma}}{dE_\gamma}=\int_0^\infty dE_0 \,  W_{2\gamma} \, G(E_\gamma, \xi /\omega, E_0),
\end{equation}
with
\begin{equation}
W_{2\gamma} = \frac{1}{(E^{max}_\gamma-E_\gamma^{min})}\Theta (E_\gamma - E_\gamma^{min}) \Theta (E_\gamma^{max} - E_\gamma).
\end{equation}
In this case the energy of the photons is in the range:
\begin{equation}
E_\gamma^{min}=\frac{M_{h_\ell}}{2\gamma (1 + \beta)} \leq E_\gamma \leq  \frac{M_{h_\ell}}{2\gamma (1 - \beta)} =E_\gamma^{max},
\end{equation}
where 
\begin{eqnarray}
\gamma &=& \frac{E_{h_\ell}}{M_{h_\ell}}, \ \beta = \sqrt{1 - \frac{M_{h_\ell}^2}{E_{h_\ell}^2}},   \nonumber 
\end{eqnarray}
and
\begin{eqnarray}
E_{h_\ell}&=& \frac{1}{ 4 M_\chi} \left( M_{h_\ell}^2 + 4 M_\chi^2 - M_{Z_\ell}^2\right).
\end{eqnarray}
The photon flux for this channel is suppressed by the small branching ratios of $h_\ell \to \gamma \gamma$. The photon flux produced by the final state radiation can be written as
\begin{equation}
\frac{d\Phi_{FSR}}{dE_\gamma}= \frac{1}{8 \pi M_\chi^2} \frac{ d ( \sigma v_\text{rel} 
(\chi \chi \to q  \bar{q} \gamma))}{dE_\gamma}J_\text{ann},
\end{equation}
As we discussed before, the photon flux produced by the final state radiation processes is highly suppressed. 

In  Fig.~\ref{EnSpec} we show the contribution of the different processes to the total photon flux from dark matter annihilation. Notice that the transition from the continuum to the gamma lines can easily be seen. To illustrate this transition, we choose a benchmark scenario using as input parameters $M_\chi=36$ TeV and $M_{Z_\ell}=16.5$ TeV that satisfies the relic density, direct detection and $N_{eff}$ bounds. 
\end{itemize}
\begin{itemize}
\item{Neutrino Signatures: $\chi \chi \to Z_\ell Z_\ell \to \nu_i \bar{\nu}_i \nu_i \bar{\nu}_i$.}
\begin{figure} [t]  
        \centering       
        \includegraphics[width=0.45\textwidth]{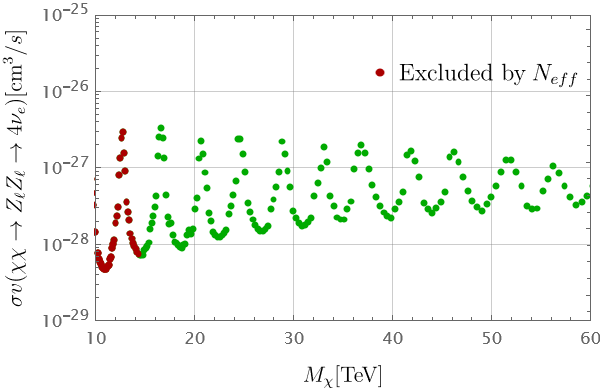}
         \caption{Cross section for $\chi \chi \to Z_\ell Z_\ell \to 4\nu_e$.}
         \label{INDNU}
\end{figure}
In this theory the dark matter annihilation into two neutrinos mediated by the leptophilic gauge boson $Z_\ell$ is velocity suppressed. However, one can have channels with four neutrinos when the dark matter candidate annihilates into two of the gauge bosons and then the leptophilic gauge boson decays into two neutrinos. 
In Fig.~\ref{INDNU} we show the predictions for the annihilation cross section for $\chi \chi \to Z_\ell Z_\ell \to 4\nu_e$ as a function of the dark matter mass.
Unfortunately, the predictions for these channels are far from the experimental limits. See Ref.~\cite{IceCube:2023ies} for the experimental bounds on neutrino lines from dark matter annihilation.
\end{itemize}
\section{MAJORANA NEUTRINOS}
\label{Majorana}
The SM neutrinos could be Majorana or Dirac neutrinos. In the previous sections, we discussed the predictions in the minimal model based on $U(1)_\ell$ where neutrinos are predicted to be Dirac fermions. In this theory, one can generate Majorana neutrino masses implementing the canonical seesaw mechanism in a simple way adding a new Higgs boson, $\phi \sim ({\bf{1}},{\bf{1}},0,-2)$, and using the following Yukawa interactions:
\begin{equation}
- {\cal L} \supset Y_\nu \bar{\ell}_L i \sigma_2 H^* \nu_R + \lambda_R \nu_R^T C \nu_R \phi \ + \ {\rm h.c.}.
\end{equation}
Therefore, in the Majorana case we have three scalar fields, $H$, $S$ and $\phi$. The  most general scalar potential in this theory can be written as
 \begin{widetext}
 \begin{eqnarray}
 V(H,S,\phi)&=&-m_H^2 H ^{\dagger}H+\lambda(H^{\dagger}H)^2-m_s^2 S ^{\dagger}S 
 + \lambda_s (S^{\dagger}S)^2-m_{\phi}^2 \phi ^{\dagger}\phi 
 + \lambda_{\phi}(\phi^{\dagger}\phi)^2 \nonumber \\
 &+& \lambda_1(H^{\dagger}H)S^{\dagger}S + \lambda_2(H^{\dagger}H)\phi^{\dagger}\phi 
 + \lambda_3(S^{\dagger}S)\phi^{\dagger}\phi.
 \end{eqnarray}
 \end{widetext}
The new scalar field in this theory can be written as 
\begin{eqnarray}
\phi &=& \frac{1}{\sqrt{2}}\left( v_{\phi} + h_{\phi} \right) e^{i \sigma_{\phi}/v_{\phi}} .
\end{eqnarray} 
This scalar potential has the global symmetry: 
$O(4)_H \otimes U(1)_\phi \otimes U(1)_S $. 
In our notation, the physical CP-even Higgses are: $h, H_1$, and $H_2$. They are related to the unphysical fields by the relation: $(h_0,h_S,h_\phi)^T= (h,H_1,H_2)^T U^T$.
There are three CP-odd Higgses and two of them are Goldstone's bosons eaten by the neutral gauge bosons. The gauge symmetry of this theory allows a seven-dimensional term in the potential:
\begin{eqnarray}
 V(H,S,\phi) \supset  - \lambda_{M} \frac{S^4 \phi^3}{\Lambda^3}+ {\rm h.c.}. 
 \end{eqnarray}
 This term breaks the $U(1)_\phi \otimes U(1)_S$ global symmetry of the potential and one gets a pseudo-Nambu-Goldstone boson, the Majoron $J$~\cite{Chikashige:1980ui}. The mass matrix for the new CP-odd Higgses in the basis, $(\sigma_s, \sigma_\phi)$, is given by
\begin{equation}
 {M_{odd}^2=\frac{\lambda_M v_S^2 v_\phi}{2 \sqrt{2}\Lambda^3}\begin{pmatrix}
16 v_\phi^2 & 12 v_S v_\phi \\
12 v_S v_\phi & 9 v_s^2 \\
\end{pmatrix}.}
\end{equation}
Therefore, one obtains a massless Goldstone's boson, $G_\ell$, and a massive pseudo-Goldstone's boson with mass given by
\begin{equation}
M_J^2= \frac{\lambda_M}{2\sqrt{2} \Lambda^3} v_S^2 v_\phi \left( 16 v_\phi^2 + 9 v_S^2 \right).
\end{equation}
The CP-odd Higgs eigenstates are defined by
\begin{eqnarray}
\begin{pmatrix}
\sigma_s \\
\sigma_{\phi} 
\end{pmatrix} =
\begin{pmatrix}
\cos \beta & \sin \beta \\
- \sin \beta & \cos \beta \\
\end{pmatrix}
\begin{pmatrix}
G_{\ell} \\
J
\end{pmatrix},
\end{eqnarray}
where 
\begin{equation}
\tan 2 \beta= \frac{24 v_S v_\phi}{16 v_\phi^2- 9 v_S^2}.
\end{equation}
 In this theory, the mass of the new gauge boson associated to lepton number is given by
 \begin{equation}
 M_{Z_\ell}^2=g_\ell^2 \left( \frac{9}{4} v_S^2 + 4 v_\phi^2 \right).
 \end{equation}
 Using $v_S=2 v \sin \beta/3$ and $v_\phi=v \cos \beta/2$ one can write $M_{Z_\ell}=g_\ell v$.
The SM neutrino masses are generated through the type I seesaw mechanism and the neutrino mass matrix is given by
\begin{equation}
    M_\nu = \frac{v_0^2}{2} Y_\nu M_N^{-1} Y_\nu^T,
\end{equation}
where 
\begin{equation}
M_N=\sqrt{2} \lambda_R v_\phi= \frac{\lambda_R}{\sqrt{2}} \frac{M_{Z_\ell}}{g_\ell} \sin \beta.
\label{MNZL}
\end{equation}
Notice that the seesaw mechanism has to be realized at the low scale once we use the upper bound on the ratio $M_{Z_\ell}/g_\ell$ coming from the cosmological bounds on the dark matter relic density and the perturbative bound on the Yukawa coupling $\lambda_R$, i.e. $\lambda^{ij}_R \leq 2 \sqrt{\pi}$. 
\begin{figure} [h]  
        \centering            \includegraphics[width=0.55\textwidth]{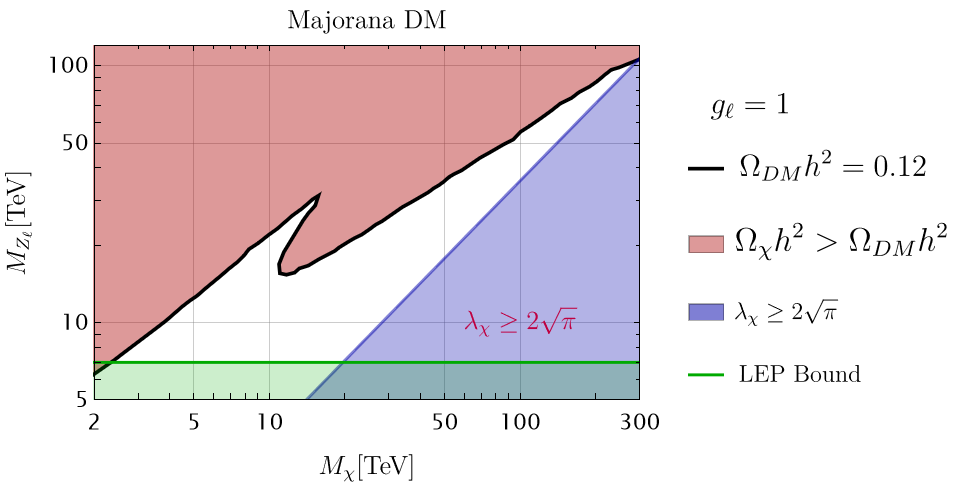}
         \caption{ Allowed parameter space by the dark matter relic density and perturbative bound in the $M_{Z_\ell}$ - $M_{\chi}$ plane when $g_{\ell}=1.0$ , $M_{H_i} = 200$ GeV and the mixing between new Higgses ($H_1$ and $H_2$) is $\pi/6$. Here we consider the Sommerfeld enhancement mediated by the two new Higgses. The black line represents the relic density $\Omega_{DM} h^2 = 0.12$. The light red-shaded region is excluded by the overproduction of dark matter and the blue-shaded region is excluded by the perturbative bound on $\lambda_\chi$. The green line shows the collider bounds from LEP.}
         \label{RelicMaj}
\end{figure}
\begin{itemize}
\item{\textit{Relic density:}} In the case of Majorana neutrinos our dark matter candidate has several new annihilation channels:
\begin{equation}
\chi \chi \to JJ, Z_\ell J, H_i H_j, Z_\ell H_i, J H_i, NN. 
\end{equation}
See the Feynman graphs in Appendix~\ref{Annihilation-Dirac}. 
It is important to mention that some of the annihilation channels present in the case of the Dirac neutrinos have extra contributions mediated by the new Higgses. In particular, the new annihilation channels $\chi \chi \to J Z_\ell, Z_\ell H_i$ play an important role to achieve the correct relic density. Here we include the Sommerfeld enhancement due to the presence of the new two Higgses.

In Fig.~\ref{RelicMaj} we show the allowed parameter space by the dark matter relic density and perturbative bound in the $M_{Z_\ell}$ - $M_{\chi}$ plane when $g_{\ell}=1.0 $ and $M_{H_i} = 200$ GeV. Here we consider the Sommerfeld enhancement mediated by the two new Higgses. The black line represents the relic density $\Omega_{DM} h^2 = 0.12$. The light red-shaded region is excluded by the overproduction of dark matter, while the purple-shaded region is excluded by the perturbative bound on $\lambda_\chi$. The green line shows the collider bounds from LEP. Notice that thanks to presence of new annihilation channels one can achieve the correct relic density even when the leptophilic gauge boson is much heavier than in the case when the neutrinos are Dirac fermions. Therefore, the upper bound on the symmetry breaking scale is much larger since $M_{Z_\ell}$ can be around $100$ TeV. 
\begin{figure} [t]  
        \centering       
        \includegraphics[width=0.45\textwidth]{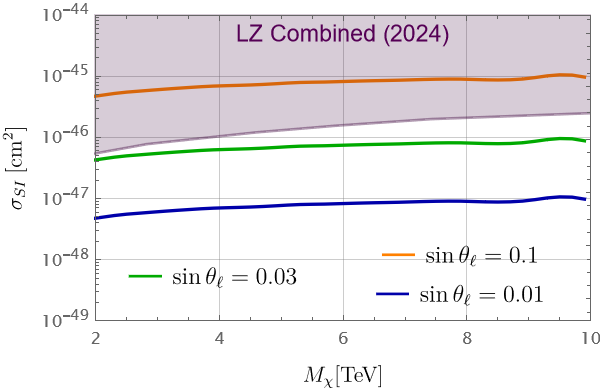}
         \caption{Spin independent dark matter-nucleon cross section as a function of dark matter mass when $M_{H_i}=200$ GeV. The purple-shaded region is excluded by the LZ experiment ~\cite{aalbers2024darkmattersearchresults}. Here we used $f_N=0.308$ ~\cite{Hoferichter_2017}.}
         \label{DDM}
\end{figure}
\item{\textit{Direct detection:}} In Fig.~\ref{DDM} we show the spin-independent dark matter-nucleon cross section as a function of dark matter mass when $M_{H_i}=200$ GeV. The red-shaded region is excluded by the LZ experiment~\cite{aalbers2024darkmattersearchresults}. Here we used $f_N=0.308$ and show the predictions for the scenarios where we achieve the correct relic density. Notice that one can satisfy the experimental bounds when the mixing angle between the SM Higgs and the new Higgses is smaller than approximately $0.03$.
\item{\textit{Gamma lines:}} Using the previous results we show in Figs.~\ref{IND2gammagamma} and \ref{IND2gammaZ} the predictions for the gamma lines. In Fig.~\ref{IND2gammagamma} we show the numerical results for the thermal average cross section for $\chi \chi \to \gamma \gamma $ as a function of dark matter mass, $M_{\chi}$. Here all the points satisfy the relic density constraints and we use $M_{H_1}=M_{H_2}=200$ GeV, $g_\ell=1.0$, $M_{\Psi^-}=1.2 M_\chi$ and $M_{\rho^-}=3 M_\chi$. The dashed line shows the projected CTA experimental bounds~\cite{CTAO:2024wvb}. 
\begin{figure} [h]  
        \centering       
        \includegraphics[width=0.45\textwidth]{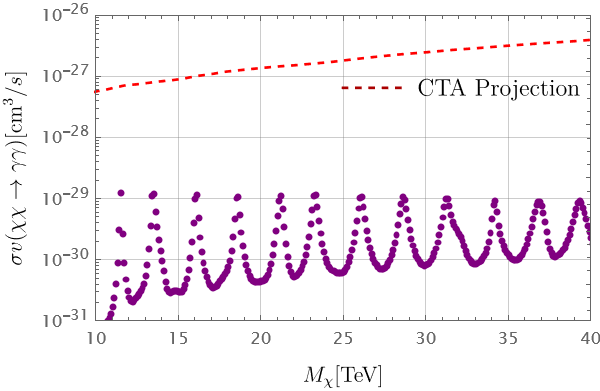}
         \caption{Cross section for $\chi \chi \to \gamma \gamma $ as a function of dark matter mass, $M_{\chi}$. Here all the points satisfy the relic density constraints. We use $M_{H_1}=M_{H_2}=200$ GeV, $g_\ell=1$ , $M_{\Psi^-}=1.2 M_\chi$ and $M_{\rho^-}=3 M_\chi$.The dashed line shows the CTA projection ~\cite{CTAO:2024wvb}. }
         \label{IND2gammagamma}
\end{figure}
\begin{figure} [h]  
        \centering       
        \includegraphics[width=0.45\textwidth]{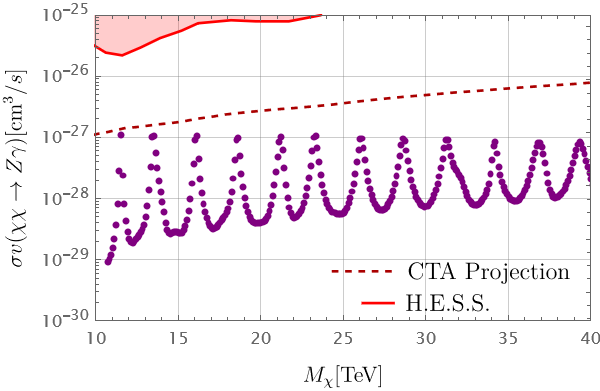}
         \caption{Cross section for $\chi \chi \to Z \gamma$ as a function of dark matter mass, $M_{\chi}$. Here all the scenarios satisfy the relic density constraints. We use $M_{H_1}=M_{H_2}=200$ GeV, $g_\ell=1.0$ , $M_{\Psi^-}=1.2 M_\chi$ and $M_{\rho^-}=3 M_\chi$.The red-shaded region is excluded by the H.E.S.S. ~\cite{Abramowski_2013} experiment and the dashed red line shows the CTA projection ~\cite{CTAO:2024wvb}. }
         \label{IND2gammaZ}
\end{figure}

In Fig.~\ref{IND2gammaZ} we show the numerical values for the cross section for $\chi \chi \to Z \gamma$ as a function of dark matter mass. Here all the scenarios satisfy the relic density constraints. We use $M_{H_1}=M_{H_2}=200$ GeV, $g_\ell=1.0$, $M_{\Psi^-}=1.2 M_\chi$ and $M_{\rho^-}=3 M_\chi$. The dashed red line shows the CTA projection~\cite{CTAO:2024wvb}. It is important to mention that in this case the right-handed neutrinos are very massive, the cosmological bounds on $N_{eff}$ are not relevant, and all scenarios shown in Figs.~\ref{IND2gammagamma} and \ref{IND2gammaZ} are allowed. 

The allowed values for the mass of the leptophilic gauge boson in agreement with the relic density bounds are much larger than in the scenarios with Dirac neutrinos. Therefore, the cross sections for the gamma lines are much smaller. Only in a few scenarios the numerical values for the $\sigma (\chi \chi \to Z \gamma)$ are close to the projected CTA bounds and there is hope to test these predictions in the future.
\end{itemize}
\section{SUMMARY}
\label{Summary}
The origin of neutrino masses is one of the most pressing issues in particle physics. In this paper,
we discussed the minimal theory based on local lepton number for neutrino masses predicting a dark matter from anomaly cancellation. In this context, the dark matter candidate is a Majorana fermion and the neutrinos are Dirac fermions. 
We discussed in detail the properties of the 
dark matter candidate showing the relic density and direct detection constraints. We discussed the impact of the $N_{eff}$ cosmological bounds, which rule out some scenarios. We have shown that the projected SPT and SO experimental bounds~\cite{Ade_2019,Benson_2014} could test or rule out the predictions even if the ratio, $M_{Z_\ell}/g_\ell$, is above 30 TeV, far from the reach of any proposed collider experiment.
\begin{figure} [h]  
        \centering       
        \includegraphics[width=0.50\textwidth]{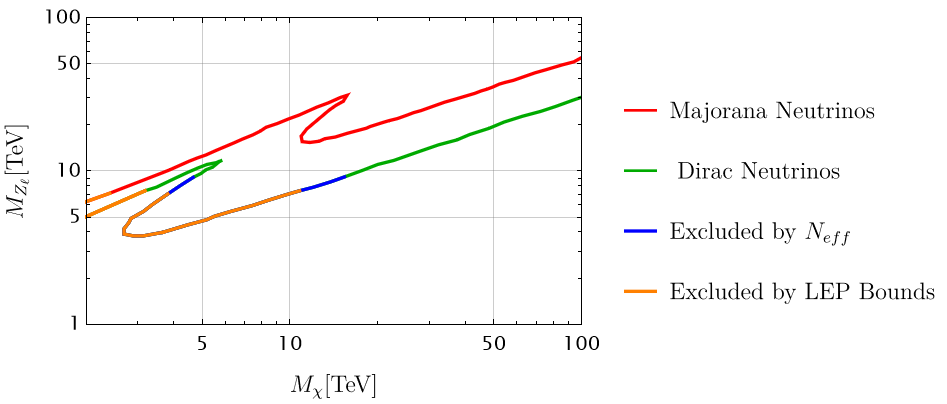}
         \caption{Allowed parameter space by the relic density bound,i.e. $\Omega_{DM} h^2=0.12$. In green, we show the results when the neutrinos are Dirac fermions, while in red we show the results when the neutrinos are Majorana fermions. In orange, we show the region excluded by the LEP bounds, while in blue we show the scenarios ruled out by the $N_{eff}$ bounds.}
         \label{COM}
\end{figure}
This theory predicts neutrino and gamma lines from dark matter annihilation. In particular, the predictions for the gamma line from the dark matter annihilation channel, $\chi \chi \to Z \gamma$, are close to the projected CTA experimental limits. 
Therefore, one can hope to test these predictions in the near future.
We computed all relevant processes contributing to the continuum photon spectrum and showed that the visibility of the gamma lines is not spoiled and one can easily see the main features of these gamma lines.

We studied the implementation of the canonical seesaw mechanism for Majorana neutrinos in this context. In this case, our dark matter candidate has several new annihilation channels. Therefore, the allowed parameter space in agreement with the relic density constraints is quite different. In this case one has extra contributions to the Sommerfeld enhancement due to the presence of the new two Higges. This theory predicts a new pseudo-Goldstone boson which couples at tree-level to neutrinos, the dark matter and the new charged fermions. This new scalar field changes significaly the predictions for dark matter annihilation and the parameter space allowed by cosmology.

In Fig.~\ref{COM} we show the allowed parameter space in the $M_{Z_\ell}-M_\chi$ plane where we obtain $\Omega_{DM} h^2=0.12$. As one can appreciate, the symmetry breaking scale can be much larger in the Majorana case. However, these simple theories for neutrino masses and dark matter could be tested in the near future and tell us that one can have a strong relation between the origin of neutrino masses and the properties of the dark matter in the universe.

{\textit{Acknowledgements:}} P.F.P. thanks the SIMONS Foundation for financial support and the
Galileo Galilei Institute for Theoretical Physics in Florence, Italy, for hospitality. This work made use of the High Performance Computing Resource
in the Core Facility for Advanced Research Computing at Case Western Reserve University.

\appendix
\section{FEYNMAN RULES}
DIRAC NEUTRINOS
\begin{eqnarray}
Z_\ell \bar{\nu}_i \nu_i &:& -i g_\ell \gamma^\mu, \\
Z_\ell \bar{e}_i e_i &:& -i g_\ell \gamma^\mu, \\
\chi \chi Z_\ell &:&  i \frac{3}{4} g_\ell \gamma^\mu \gamma^5, \\
\chi \chi h &:&  i \frac{3}{2} \frac{g_\ell M_\chi}{M_{Z_\ell}} \sin \theta_\ell, \\
\chi \chi h_\ell &:& -i \frac{3}{2} \frac{g_\ell M_\chi}{M_{Z_\ell}} \cos \theta_\ell, \\
h_\ell Z_\ell Z_\ell &:&  i {3} g_\ell M_{Z_\ell} \cos \theta_\ell \ g^{\mu \nu}, \\
h Z_\ell Z_\ell &:& - i 3 g_\ell M_{Z_\ell} \sin \theta_\ell \ g^{\mu \nu}, \\
h Z Z & :& 2i  \frac{M_{Z}^2}{v_0} \cos \theta_\ell \ g^{\mu \nu}, \\
 h_\ell Z Z & : & 2i \frac{M_{Z}^2}{v_0} \sin \theta_\ell \ g^{\mu \nu}, \\
 h W W & : & 2i   \frac{M_{W}^2}{v_0} \cos \theta_\ell \ g^{\mu \nu}, \\ 
 h_\ell W W & : & 2i   \frac{M_{W}^2}{v_0} \sin \theta_\ell \ g^{\mu \nu}, \\
 {\overline{\rho^-}} \rho^- Z_\ell &:& {-i \frac{3}{4} g_\ell \gamma^\mu \gamma^5},\\
      \overline{\Psi^-} \Psi^- Z_\ell  &:& i \frac{3}{4} g_\ell  \gamma^\mu \gamma^5, \\
      {\overline{\rho^-}} \rho^- h_\ell &:& i \frac{3}{2} \frac{g_\ell M_{\rho^-}}{M_{Z_\ell}}  \cos \theta_\ell,\\
      {\overline{\Psi^-}} \Psi^- h_\ell &:& i \frac{3}{2} \frac{g_\ell M_{\Psi^-}}{M_{Z_\ell}} 
      \cos \theta_\ell.
\end{eqnarray}
MAJORANA NEUTRINOS
\begin{eqnarray}
Z_\ell^\mu \nu \nu &:& -i g_\ell \gamma^\mu \gamma^5,\\ 
Z_\ell^\mu N N &:& i g_\ell \gamma^\mu \gamma^5\\
\chi \chi H_i &:& i \frac{3 M_\chi g_\ell}{2 M_{Z_\ell}} \frac{U_{2i}}{\sin{\beta}}, \\
N N H_i  &:&  i \frac{2 g_\ell M_N }{M_{Z_\ell}\cos{\beta}} U_{3i},  \\
\chi(p_1) \chi(p_2) J(p)   &:&  i \frac{3 g_\ell  }{2 M_{Z_\ell}} p_{\mu} \gamma^\mu\gamma^5, \\
N(p_1) N(p_2) J(p)   &:&  i \frac{ g_\ell  }{ M_{Z_\ell}} p_{\mu} \gamma^\mu\gamma^5, \\
Z_\ell Z_\ell H_i   &:&  i g_\ell M_{Z_\ell} \left( 3 \sin{\beta} U_{2i} + 4 \cos{\beta} U_{3i}\right) g^{\mu \nu}. \nonumber \\
\end{eqnarray}
\begin{widetext}
\section{DARK MATTER ANNIHILATION CHANNELS} 
\label{Annihilation-Dirac}
\begin{figure}[h]
DIRAC NEUTRINOS
\begin{eqnarray*}
\begin{gathered}
\begin{tikzpicture}[line width=1.5 pt,node distance=1 cm and 1.5 cm]
\coordinate[label = left: $\chi$] (i1);
\coordinate[below right = 1cm of i1](v1);
\coordinate[below left = 1cm of v1, label= left:$\chi$](i2);
\coordinate[above right = 1cm of v1, label=right: $\ell$] (f1);
\coordinate[below right =  1cm of v1,label=right: $\bar{\ell}$] (f2);
\draw[fermionnoarrow] (i1) -- (v1);
\draw[fermionnoarrow] (i2) -- (v1);
\draw[fermion] (v1) -- (f1);
\draw[fermion] (f2) -- (v1);
\draw[fill=gray] (v1) circle (.3cm);
\end{tikzpicture}
\end{gathered} 
&=&
\begin{gathered}
\begin{tikzpicture}[line width=1.5 pt,node distance=1 cm and 1.5 cm]
\coordinate[label =left: $\chi$] (i1);
\coordinate[below right= 1cm of i1](v1);
\coordinate[ right= 0.5cm of v1,label= above:$Z_\ell$](vaux);
\coordinate[below left= 1cm of v1, label= left: $\chi$](i2);
\coordinate[right = 1 cm of v1](v2);
\coordinate[above right = 1 cm of v2, label=right: $e_i^- \text{,} \nu_i$] (f1);
\coordinate[below right =  1 cm of v2,label=right: $e_i^+ \text{,} \bar{\nu_i}$] (f2);
\draw[fermionnoarrow] (i1) -- (v1);
\draw[fermionnoarrow] (i2) -- (v1);
\draw[vector] (v1) -- (v2);
\draw[fermion] (v2) -- (f1);
\draw[fermion] (f2) -- (v2);
\draw[fill=cyan] (v1) circle (.1cm);
\draw[fill=cyan] (v2) circle (.1cm);
\end{tikzpicture}
\end{gathered}
\quad   \! +
\begin{gathered}
\begin{tikzpicture}[line width=1.5 pt,node distance=1 cm and 1.5 cm]
\coordinate[label =left: $\chi$] (i1);
\coordinate[below right= 1cm of i1](v1);
\coordinate[ right= 0.5cm of v1,label= above:$h_\ell \text{,} h$](vaux);
\coordinate[below left= 1cm of v1, label= left: $\chi$](i2);
\coordinate[right = 1 cm of v1](v2);
\coordinate[above right = 1 cm of v2, label=right: $e_i^- \text{,} \nu_i$] (f1);
\coordinate[below right =  1 cm of v2,label=right: $e_i^+ \text{,} \bar{\nu_i}$] (f2);
\draw[fermionnoarrow] (i1) -- (v1);
\draw[fermionnoarrow] (i2) -- (v1);
\draw[scalarnoarrow] (v1) -- (v2);
\draw[fermion] (v2) -- (f1);
\draw[fermion] (f2) -- (v2);
\draw[fill=black] (v1) circle (.1cm);
\draw[fill=black] (v2) circle (.1cm);
\end{tikzpicture}
\end{gathered}\\
\begin{gathered}
\begin{tikzpicture}[line width=1.5 pt,node distance=1 cm and 1.5 cm]
\coordinate[label = left: $\chi$] (i1);
\coordinate[below right = 1cm of i1](v1);
\coordinate[below left = 1cm of v1, label= left:$\chi$](i2);
\coordinate[above right = 1cm of v1, label=right: $Z_\ell$] (f1);
\coordinate[below right =  1cm of v1,label=right: $Z_\ell$] (f2);
\draw[fermionnoarrow] (i1) -- (v1);
\draw[fermionnoarrow] (i2) -- (v1);
\draw[vector] (v1) -- (f1);
\draw[vector] (f2) -- (v1);
\draw[fill=gray] (v1) circle (.3cm);
\end{tikzpicture}
\end{gathered} 
&=&
\begin{gathered}
\begin{tikzpicture}[line width=1.5 pt,node distance=1 cm and 1.5 cm]
\coordinate[label =left: $\chi$] (i1);
\coordinate[right= 1cm of i1](v1);
\coordinate[below= 0.5cm of v1](vaux);
\coordinate[right = 1cm of v1, label= right:$Z_\ell$](f1);
\coordinate[below = 1 cm of v1](v2);
\coordinate[left = 1 cm of v2, label=left: $\chi$] (i2);
\coordinate[right =  1 cm of v2,label=right: $Z_\ell$] (f2);
\draw[fermionnoarrow] (i1) -- (v1);
\draw[fermionnoarrow] (v1) -- (v2);
\draw[fermionnoarrow] (i2) -- (v2);
\draw[vector] (f2) -- (v2);
\draw[vector] (f1) -- (v1);
\draw[fill=cyan] (v1) circle (.1cm);
\draw[fill=cyan] (v2) circle (.1cm);
\end{tikzpicture}
\end{gathered}
\quad   \! +
\begin{gathered}
\begin{tikzpicture}[line width=1.5 pt,node distance=1 cm and 1.5 cm]
\coordinate[label =left: $\chi$] (i1);
\coordinate[right= 1cm of i1](v1);
\coordinate[below= 0.5cm of v1,label](vaux);
\coordinate[right = 1cm of v1, label= right:$Z_\ell$](f1);
\coordinate[below = 1 cm of v1](v2);
\coordinate[left = 1 cm of v2, label=left: $\chi$] (i2);
\coordinate[right =  1 cm of v2,label=right: $Z_\ell$] (f2);
\draw[fermionnoarrow] (i1) -- (v1);
\draw[fermionnoarrow] (v1) -- (v2);
\draw[fermionnoarrow] (i2) -- (v2);
\draw[vector] (f2) -- (v1);
\draw[vector] (f1) -- (v2);
\draw[fill=cyan] (v1) circle (.1cm);
\draw[fill=cyan] (v2) circle (.1cm);
\end{tikzpicture}
\end{gathered}
+
\begin{gathered}
\begin{tikzpicture}[line width=1.5 pt,node distance=1 cm and 1.5 cm]
\coordinate[label =left: $\chi$] (i1);
\coordinate[below right= 1cm of i1](v1);
\coordinate[ right= 0.5cm of v1,label= above:$h_\ell \text{,} h$](vaux);
\coordinate[below left= 1cm of v1, label= left: $\chi$](i2);
\coordinate[right = 1 cm of v1](v2);
\coordinate[above right = 1 cm of v2, label=right: $Z_\ell$] (f1);
\coordinate[below right =  1 cm of v2,label=right: $Z_\ell$] (f2);
\draw[fermionnoarrow] (i1) -- (v1);
\draw[fermionnoarrow] (i2) -- (v1);
\draw[scalarnoarrow] (v1) -- (v2);
\draw[vector] (v2) -- (f1);
\draw[vector] (f2) -- (v2);
\draw[fill=black] (v1) circle (.1cm);
\draw[fill=cyan] (v2) circle (.1cm);
\end{tikzpicture}
\end{gathered} \\
\begin{gathered}
\begin{tikzpicture}[line width=1.5 pt,node distance=1 cm and 1.5 cm]
\coordinate[label = left: $\chi$] (i1);
\coordinate[below right = 1cm of i1](v1);
\coordinate[below left = 1cm of v1, label= left:$\chi$](i2);
\coordinate[above right = 1cm of v1, label=right: $Z_\ell$] (f1);
\coordinate[below right =  1cm of v1,label=right: $h_\ell(h)$] (f2);
\draw[fermionnoarrow] (i1) -- (v1);
\draw[fermionnoarrow] (i2) -- (v1);
\draw[vector] (v1) -- (f1);
\draw[scalarnoarrow] (f2) -- (v1);
\draw[fill=gray] (v1) circle (.3cm);
\end{tikzpicture}
\end{gathered} 
&=&
\begin{gathered}
\begin{tikzpicture}[line width=1.5 pt,node distance=1 cm and 1.5 cm]
\coordinate[label =left: $\chi$] (i1);
\coordinate[right= 1cm of i1](v1);
\coordinate[below= 0.5cm of v1](vaux);
\coordinate[right = 1cm of v1, label= right:$Z_\ell$](f1);
\coordinate[below = 1 cm of v1](v2);
\coordinate[left = 1 cm of v2, label=left: $\chi$] (i2);
\coordinate[right =  1 cm of v2,label=right: $h_\ell(h)$] (f2);
\draw[fermionnoarrow] (i1) -- (v1);
\draw[fermionnoarrow] (v1) -- (v2);
\draw[fermionnoarrow] (i2) -- (v2);
\draw[scalarnoarrow] (f2) -- (v2);
\draw[vector] (f1) -- (v1);
\draw[fill=cyan] (v1) circle (.1cm);
\draw[fill=black] (v2) circle (.1cm);
\end{tikzpicture}
\end{gathered}
\quad   \! +
\begin{gathered}
\begin{tikzpicture}[line width=1.5 pt,node distance=1 cm and 1.5 cm]
\coordinate[label =left: $\chi$] (i1);
\coordinate[right= 1cm of i1](v1);
\coordinate[below= 0.5cm of v1,label](vaux);
\coordinate[right = 1cm of v1, label= right:$Z_\ell$](f1);
\coordinate[below = 1 cm of v1](v2);
\coordinate[left = 1 cm of v2, label=left: $\chi$] (i2);
\coordinate[right =  1 cm of v2,label=right: $h_\ell(h)$] (f2);
\draw[fermionnoarrow] (i1) -- (v1);
\draw[fermionnoarrow] (v1) -- (v2);
\draw[fermionnoarrow] (i2) -- (v2);
\draw[scalarnoarrow] (f2) -- (v1);
\draw[vector] (f1) -- (v2);
\draw[fill=black] (v1) circle (.1cm);
\draw[fill=cyan] (v2) circle (.1cm);
\end{tikzpicture}
\end{gathered}
+
\begin{gathered}
\begin{tikzpicture}[line width=1.5 pt,node distance=1 cm and 1.5 cm]
\coordinate[label =left: $\chi$] (i1);
\coordinate[below right= 1cm of i1](v1);
\coordinate[ right= 0.5cm of v1,label= above:$Z_\ell$](vaux);
\coordinate[below left= 1cm of v1, label= left: $\chi$](i2);
\coordinate[right = 1 cm of v1](v2);
\coordinate[above right = 1 cm of v2, label=right: $Z_\ell$] (f1);
\coordinate[below right =  1 cm of v2,label=right: $h_\ell(h)$] (f2);
\draw[fermionnoarrow] (i1) -- (v1);
\draw[fermionnoarrow] (i2) -- (v1);
\draw[vector] (v1) -- (v2);
\draw[vector] (v2) -- (f1);
\draw[scalarnoarrow] (f2) -- (v2);
\draw[fill=black] (v1) circle (.1cm);
\draw[fill=cyan] (v2) circle (.1cm);
\end{tikzpicture}
\end{gathered} \\
\begin{gathered}
\begin{tikzpicture}[line width=1.5 pt,node distance=1 cm and 1.5 cm]
\coordinate[label = left: $\chi$] (i1);
\coordinate[below right = 1cm of i1](v1);
\coordinate[below left = 1cm of v1, label= left:$\chi$](i2);
\coordinate[above right = 1cm of v1, label=right: $h_\ell$] (f1);
\coordinate[below right =  1cm of v1,label=right: $h_\ell$] (f2);
\draw[fermionnoarrow] (i1) -- (v1);
\draw[fermionnoarrow] (i2) -- (v1);
\draw[scalarnoarrow] (v1) -- (f1);
\draw[scalarnoarrow] (f2) -- (v1);
\draw[fill=gray] (v1) circle (.3cm);
\end{tikzpicture}
\end{gathered} 
&=&
\begin{gathered}
\begin{tikzpicture}[line width=1.5 pt,node distance=1 cm and 1.5 cm]
\coordinate[label =left: $\chi$] (i1);
\coordinate[right= 1cm of i1](v1);
\coordinate[below= 0.5cm of v1](vaux);
\coordinate[right = 1cm of v1, label= right:$h_\ell$](f1);
\coordinate[below = 1 cm of v1](v2);
\coordinate[left = 1 cm of v2, label=left: $\chi$] (i2);
\coordinate[right =  1 cm of v2,label=right: $h_\ell$] (f2);
\draw[fermionnoarrow] (i1) -- (v1);
\draw[fermionnoarrow] (v1) -- (v2);
\draw[fermionnoarrow] (i2) -- (v2);
\draw[scalarnoarrow] (f2) -- (v2);
\draw[scalarnoarrow] (f1) -- (v1);
\draw[fill=black] (v1) circle (.1cm);
\draw[fill=black] (v2) circle (.1cm);
\end{tikzpicture}
\end{gathered}
\quad   \! +
\begin{gathered}
\begin{tikzpicture}[line width=1.5 pt,node distance=1 cm and 1.5 cm]
\coordinate[label =left: $\chi$] (i1);
\coordinate[right= 1cm of i1](v1);
\coordinate[below= 0.5cm of v1,label](vaux);
\coordinate[right = 1cm of v1, label= right:$h_\ell$](f1);
\coordinate[below = 1 cm of v1](v2);
\coordinate[left = 1 cm of v2, label=left: $\chi$] (i2);
\coordinate[right =  1 cm of v2,label=right: $h_\ell$] (f2);
\draw[fermionnoarrow] (i1) -- (v1);
\draw[fermionnoarrow] (v1) -- (v2);
\draw[fermionnoarrow] (i2) -- (v2);
\draw[scalarnoarrow] (f2) -- (v1);
\draw[scalarnoarrow] (f1) -- (v2);
\draw[fill=black] (v1) circle (.1cm);
\draw[fill=black] (v2) circle (.1cm);
\end{tikzpicture}
\end{gathered}
+
\begin{gathered}
\begin{tikzpicture}[line width=1.5 pt,node distance=1 cm and 1.5 cm]
\coordinate[label =left: $\chi$] (i1);
\coordinate[below right= 1cm of i1](v1);
\coordinate[ right= 0.5cm of v1,label= above:$h_\ell\text{,}h$](vaux);
\coordinate[below left= 1cm of v1, label= left: $\chi$](i2);
\coordinate[right = 1 cm of v1](v2);
\coordinate[above right = 1 cm of v2, label=right: $h_\ell$] (f1);
\coordinate[below right =  1 cm of v2,label=right: $h_\ell$] (f2);
\draw[fermionnoarrow] (i1) -- (v1);
\draw[fermionnoarrow] (i2) -- (v1);
\draw[scalarnoarrow] (v1) -- (v2);
\draw[scalarnoarrow] (v2) -- (f1);
\draw[scalarnoarrow] (f2) -- (v2);
\draw[fill=black] (v1) circle (.1cm);
\draw[fill=black] (v2) circle (.1cm);
\end{tikzpicture}
\end{gathered}\\
\begin{gathered}
\begin{tikzpicture}[line width=1.5 pt,node distance=1 cm and 1.5 cm]
\coordinate[label = left: $\chi$] (i1);
\coordinate[below right = 1cm of i1](v1);
\coordinate[below left = 1cm of v1, label= left:$\chi$](i2);
\coordinate[above right = 1cm of v1, label=right: $h$] (f1);
\coordinate[below right =  1cm of v1,label=right: $h$] (f2);
\draw[fermionnoarrow] (i1) -- (v1);
\draw[fermionnoarrow] (i2) -- (v1);
\draw[scalarnoarrow] (v1) -- (f1);
\draw[scalarnoarrow] (f2) -- (v1);
\draw[fill=gray] (v1) circle (.3cm);
\end{tikzpicture}
\end{gathered} 
&=&
\begin{gathered}
\begin{tikzpicture}[line width=1.5 pt,node distance=1 cm and 1.5 cm]
\coordinate[label =left: $\chi$] (i1);
\coordinate[right= 1cm of i1](v1);
\coordinate[below= 0.5cm of v1](vaux);
\coordinate[right = 1cm of v1, label= right:$h$](f1);
\coordinate[below = 1 cm of v1](v2);
\coordinate[left = 1 cm of v2, label=left: $\chi$] (i2);
\coordinate[right =  1 cm of v2,label=right: $h$] (f2);
\draw[fermionnoarrow] (i1) -- (v1);
\draw[fermionnoarrow] (v1) -- (v2);
\draw[fermionnoarrow] (i2) -- (v2);
\draw[scalarnoarrow] (f2) -- (v2);
\draw[scalarnoarrow] (f1) -- (v1);
\draw[fill=black] (v1) circle (.1cm);
\draw[fill=black] (v2) circle (.1cm);
\end{tikzpicture}
\end{gathered}
\quad   \! +
\begin{gathered}
\begin{tikzpicture}[line width=1.5 pt,node distance=1 cm and 1.5 cm]
\coordinate[label =left: $\chi$] (i1);
\coordinate[right= 1cm of i1](v1);
\coordinate[below= 0.5cm of v1,label](vaux);
\coordinate[right = 1cm of v1, label= right:$h$](f1);
\coordinate[below = 1 cm of v1](v2);
\coordinate[left = 1 cm of v2, label=left: $\chi$] (i2);
\coordinate[right =  1 cm of v2,label=right: $h$] (f2);
\draw[fermionnoarrow] (i1) -- (v1);
\draw[fermionnoarrow] (v1) -- (v2);
\draw[fermionnoarrow] (i2) -- (v2);
\draw[scalarnoarrow] (f2) -- (v1);
\draw[scalarnoarrow] (f1) -- (v2);
\draw[fill=black] (v1) circle (.1cm);
\draw[fill=black] (v2) circle (.1cm);
\end{tikzpicture}
\end{gathered}
+
\begin{gathered}
\begin{tikzpicture}[line width=1.5 pt,node distance=1 cm and 1.5 cm]
\coordinate[label =left: $\chi$] (i1);
\coordinate[below right= 1cm of i1](v1);
\coordinate[ right= 0.5cm of v1,label= above:$h_\ell \text{,}h$](vaux);
\coordinate[below left= 1cm of v1, label= left: $\chi$](i2);
\coordinate[right = 1 cm of v1](v2);
\coordinate[above right = 1 cm of v2, label=right: $h$] (f1);
\coordinate[below right =  1 cm of v2,label=right: $h$] (f2);
\draw[fermionnoarrow] (i1) -- (v1);
\draw[fermionnoarrow] (i2) -- (v1);
\draw[scalarnoarrow] (v1) -- (v2);
\draw[scalarnoarrow] (v2) -- (f1);
\draw[scalarnoarrow] (f2) -- (v2);
\draw[fill=black] (v1) circle (.1cm);
\draw[fill=black] (v2) circle (.1cm);
\end{tikzpicture}
\end{gathered}\\
\begin{gathered}
\begin{tikzpicture}[line width=1.5 pt,node distance=1 cm and 1.5 cm]
\coordinate[label = left: $\chi$] (i1);
\coordinate[below right = 1cm of i1](v1);
\coordinate[below left = 1cm of v1, label= left:$\chi$](i2);
\coordinate[above right = 1cm of v1, label=right: $h_\ell
$] (f1);
\coordinate[below right =  1cm of v1,label=right: $h$] (f2);
\draw[fermionnoarrow] (i1) -- (v1);
\draw[fermionnoarrow] (i2) -- (v1);
\draw[scalarnoarrow] (v1) -- (f1);
\draw[scalarnoarrow] (f2) -- (v1);
\draw[fill=gray] (v1) circle (.3cm);
\end{tikzpicture}
\end{gathered} 
&=&
\begin{gathered}
\begin{tikzpicture}[line width=1.5 pt,node distance=1 cm and 1.5 cm]
\coordinate[label =left: $\chi$] (i1);
\coordinate[right= 1cm of i1](v1);
\coordinate[below= 0.5cm of v1](vaux);
\coordinate[right = 1cm of v1, label= right:$h_\ell$](f1);
\coordinate[below = 1 cm of v1](v2);
\coordinate[left = 1 cm of v2, label=left: $\chi$] (i2);
\coordinate[right =  1 cm of v2,label=right: $h$] (f2);
\draw[fermionnoarrow] (i1) -- (v1);
\draw[fermionnoarrow] (v1) -- (v2);
\draw[fermionnoarrow] (i2) -- (v2);
\draw[scalarnoarrow] (f2) -- (v2);
\draw[scalarnoarrow] (f1) -- (v1);
\draw[fill=black] (v1) circle (.1cm);
\draw[fill=black] (v2) circle (.1cm);
\end{tikzpicture}
\end{gathered}
\quad   \! +
\begin{gathered}
\begin{tikzpicture}[line width=1.5 pt,node distance=1 cm and 1.5 cm]
\coordinate[label =left: $\chi$] (i1);
\coordinate[right= 1cm of i1](v1);
\coordinate[below= 0.5cm of v1,label](vaux);
\coordinate[right = 1cm of v1, label= right:$h_\ell$](f1);
\coordinate[below = 1 cm of v1](v2);
\coordinate[left = 1 cm of v2, label=left: $\chi$] (i2);
\coordinate[right =  1 cm of v2,label=right: $h$] (f2);
\draw[fermionnoarrow] (i1) -- (v1);
\draw[fermionnoarrow] (v1) -- (v2);
\draw[fermionnoarrow] (i2) -- (v2);
\draw[scalarnoarrow] (f2) -- (v1);
\draw[scalarnoarrow] (f1) -- (v2);
\draw[fill=black] (v1) circle (.1cm);
\draw[fill=black] (v2) circle (.1cm);
\end{tikzpicture}
\end{gathered}
+
\begin{gathered}
\begin{tikzpicture}[line width=1.5 pt,node distance=1 cm and 1.5 cm]
\coordinate[label =left: $\chi$] (i1);
\coordinate[below right= 1cm of i1](v1);
\coordinate[ right= 0.5cm of v1,label= above:$h_\ell \text{,}h $](vaux);
\coordinate[below left= 1cm of v1, label= left: $\chi$](i2);
\coordinate[right = 1 cm of v1](v2);
\coordinate[above right = 1 cm of v2, label=right: $h_\ell$] (f1);
\coordinate[below right =  1 cm of v2,label=right: $h$] (f2);
\draw[fermionnoarrow] (i1) -- (v1);
\draw[fermionnoarrow] (i2) -- (v1);
\draw[scalarnoarrow] (v1) -- (v2);
\draw[scalarnoarrow] (v2) -- (f1);
\draw[scalarnoarrow] (f2) -- (v2);
\draw[fill=black] (v1) circle (.1cm);
\draw[fill=black] (v2) circle (.1cm);
\end{tikzpicture}
\end{gathered}\\
\begin{gathered}
\begin{tikzpicture}[line width=1.5 pt,node distance=1 cm and 1.5 cm]
\coordinate[label = left: $\chi$] (i1);
\coordinate[below right = 1cm of i1](v1);
\coordinate[below left = 1cm of v1, label= left:$\chi$](i2);
\coordinate[above right = 1cm of v1, label=right: $W^+$] (f1);
\coordinate[below right =  1cm of v1,label=right: $W^-$] (f2);
\draw[fermionnoarrow] (i1) -- (v1);
\draw[fermionnoarrow] (i2) -- (v1);
\draw[vector] (v1) -- (f1);
\draw[vector] (f2) -- (v1);
\draw[fill=gray] (v1) circle (.3cm);
\end{tikzpicture}
\end{gathered} 
&=&
\begin{gathered}
\begin{tikzpicture}[line width=1.5 pt,node distance=1 cm and 1.5 cm]
\coordinate[label =left: $\chi$] (i1);
\coordinate[below right= 1cm of i1](v1);
\coordinate[ right= 0.5cm of v1,label= above:$h_\ell \text{,}h$](vaux);
\coordinate[below left= 1cm of v1, label= left: $\chi$](i2);
\coordinate[right = 1 cm of v1](v2);
\coordinate[above right = 1 cm of v2, label=right: $W^+$] (f1);
\coordinate[below right =  1 cm of v2,label=right: $W^-$] (f2);
\draw[fermionnoarrow] (i1) -- (v1);
\draw[fermionnoarrow] (i2) -- (v1);
\draw[scalarnoarrow] (v1) -- (v2);
\draw[vector] (v2) -- (f1);
\draw[vector] (f2) -- (v2);
\draw[fill=black] (v1) circle (.1cm);
\draw[fill=red] (v2) circle (.1cm);
\end{tikzpicture}
\end{gathered} \\
\begin{gathered}
\begin{tikzpicture}[line width=1.5 pt,node distance=1 cm and 1.5 cm]
\coordinate[label = left: $\chi$] (i1);
\coordinate[below right = 1cm of i1](v1);
\coordinate[below left = 1cm of v1, label= left:$\chi$](i2);
\coordinate[above right = 1cm of v1, label=right: $Z$] (f1);
\coordinate[below right =  1cm of v1,label=right: $Z$] (f2);
\draw[fermionnoarrow] (i1) -- (v1);
\draw[fermionnoarrow] (i2) -- (v1);
\draw[vector] (v1) -- (f1);
\draw[vector] (f2) -- (v1);
\draw[fill=gray] (v1) circle (.3cm);
\end{tikzpicture}
\end{gathered} 
&=&
\begin{gathered}
\begin{tikzpicture}[line width=1.5 pt,node distance=1 cm and 1.5 cm]
\coordinate[label =left: $\chi$] (i1);
\coordinate[below right= 1cm of i1](v1);
\coordinate[ right= 0.5cm of v1,label= above:$h_\ell \text{,} h $](vaux);
\coordinate[below left= 1cm of v1, label= left: $\chi$](i2);
\coordinate[right = 1 cm of v1](v2);
\coordinate[above right = 1 cm of v2, label=right: $Z$] (f1);
\coordinate[below right =  1 cm of v2,label=right: $Z$] (f2);
\draw[fermionnoarrow] (i1) -- (v1);
\draw[fermionnoarrow] (i2) -- (v1);
\draw[scalarnoarrow] (v1) -- (v2);
\draw[vector] (v2) -- (f1);
\draw[vector] (f2) -- (v2);
\draw[fill=black] (v1) circle (.1cm);
\draw[fill=red] (v2) circle (.1cm);
\end{tikzpicture}
\end{gathered}
\end{eqnarray*}
\end{figure}
\begin{figure}
MAJORANA NEUTRINOS
\label{Annihilation-Majorana}
\begin{eqnarray*}
\begin{gathered}
\begin{tikzpicture}[line width=1.5 pt,node distance=1 cm and 1.5 cm]
\coordinate[label = left: $\chi$] (i1);
\coordinate[below right = 1cm of i1](v1);
\coordinate[below left = 1cm of v1, label= left:$\chi$](i2);
\coordinate[above right = 1cm of v1, label=right: $e_i$] (f1);
\coordinate[below right =  1cm of v1,label=right: $ e_i$] (f2);
\draw[fermionnoarrow] (i1) -- (v1);
\draw[fermionnoarrow] (i2) -- (v1);
\draw[fermion] (v1) -- (f1);
\draw[fermion] (f2) -- (v1);
\draw[fill=gray] (v1) circle (.3cm);
\end{tikzpicture}
\end{gathered} 
&=&
\begin{gathered}
\begin{tikzpicture}[line width=1.5 pt,node distance=1 cm and 1.5 cm]
\coordinate[label =left: $\chi$] (i1);
\coordinate[below right= 1cm of i1](v1);
\coordinate[ right= 0.5cm of v1,label= above:$Z_\ell$](vaux);
\coordinate[below left= 1cm of v1, label= left: $\chi$](i2);
\coordinate[right = 1 cm of v1](v2);
\coordinate[above right = 1 cm of v2, label=right: $e_i$] (f1);
\coordinate[below right =  1 cm of v2,label=right: $e_i$] (f2);
\draw[fermionnoarrow] (i1) -- (v1);
\draw[fermionnoarrow] (i2) -- (v1);
\draw[vector] (v1) -- (v2);
\draw[fermion] (v2) -- (f1);
\draw[fermion] (f2) -- (v2);
\draw[fill=cyan] (v1) circle (.1cm);
\draw[fill=cyan] (v2) circle (.1cm);
\end{tikzpicture}
\end{gathered} 
+
\begin{gathered}
\begin{tikzpicture}[line width=1.5 pt,node distance=1 cm and 1.5 cm]
\coordinate[label =left: $\chi$] (i1);
\coordinate[below right= 1cm of i1](v1);
\coordinate[ right= 0.5cm of v1,label= above:{$h$, $H_i$}](vaux);
\coordinate[below left= 1cm of v1, label= left: $\chi$](i2);
\coordinate[right = 1 cm of v1](v2);
\coordinate[above right = 1 cm of v2, label=right: $e_i$] (f1);
\coordinate[below right =  1 cm of v2,label=right: $e_i$] (f2);
\draw[fermionnoarrow] (i1) -- (v1);
\draw[fermionnoarrow] (i2) -- (v1);
\draw[scalarnoarrow] (v1) -- (v2);
\draw[fermion] (v2) -- (f1);
\draw[fermion] (f2) -- (v2);
\draw[fill=black] (v1) circle (.1cm);
\draw[fill=black] (v2) circle (.1cm);
\end{tikzpicture}
\end{gathered} \\
\begin{gathered}
\begin{tikzpicture}[line width=1.5 pt,node distance=1 cm and 1.5 cm]
\coordinate[label = left: $\chi$] (i1);
\coordinate[below right = 1cm of i1](v1);
\coordinate[below left = 1cm of v1, label= left:$\chi$](i2);
\coordinate[above right = 1cm of v1, label=right: $\nu$] (f1);
\coordinate[below right =  1cm of v1,label=right: $\nu$] (f2);
\draw[fermionnoarrow] (i1) -- (v1);
\draw[fermionnoarrow] (i2) -- (v1);
\draw[fermionnoarrow] (v1) -- (f1);
\draw[fermionnoarrow] (f2) -- (v1);
\draw[fill=gray] (v1) circle (.3cm);
\end{tikzpicture}
\end{gathered} 
&=&
\begin{gathered}
\begin{tikzpicture}[line width=1.5 pt,node distance=1 cm and 1.5 cm]
\coordinate[label =left: $\chi$] (i1);
\coordinate[below right= 1cm of i1](v1);
\coordinate[ right= 0.5cm of v1,label= above:$Z_\ell$](vaux);
\coordinate[below left= 1cm of v1, label= left: $\chi$](i2);
\coordinate[right = 1 cm of v1](v2);
\coordinate[above right = 1 cm of v2, label=right: $\nu$] (f1);
\coordinate[below right =  1 cm of v2,label=right: $\nu$] (f2);
\draw[fermionnoarrow] (i1) -- (v1);
\draw[fermionnoarrow] (i2) -- (v1);
\draw[vector] (v1) -- (v2);
\draw[fermionnoarrow] (v2) -- (f1);
\draw[fermionnoarrow] (f2) -- (v2);
\draw[fill=cyan] (v1) circle (.1cm);
\draw[fill=cyan] (v2) circle (.1cm);
\end{tikzpicture}
\end{gathered} 
+
\begin{gathered}
\begin{tikzpicture}[line width=1.5 pt,node distance=1 cm and 1.5 cm]
\coordinate[label =left: $\chi$] (i1);
\coordinate[below right= 1cm of i1](v1);
\coordinate[ right= 0.5cm of v1,label= above:{$J$}](vaux);
\coordinate[below left= 1cm of v1, label= left: $\chi$](i2);
\coordinate[right = 1 cm of v1](v2);
\coordinate[above right = 1 cm of v2, label=right: $\nu$] (f1);
\coordinate[below right =  1 cm of v2,label=right: $\nu$] (f2);
\draw[fermionnoarrow] (i1) -- (v1);
\draw[fermionnoarrow] (i2) -- (v1);
\draw[scalarnoarrow] (v1) -- (v2);
\draw[fermionnoarrow] (v2) -- (f1);
\draw[fermionnoarrow] (f2) -- (v2);
\draw[fill=black] (v1) circle (.1cm);
\draw[fill=black] (v2) circle (.1cm);
\end{tikzpicture}
\end{gathered} \\
\begin{gathered}
\begin{tikzpicture}[line width=1.5 pt,node distance=1 cm and 1.5 cm]
\coordinate[label = left: $\chi$] (i1);
\coordinate[below right = 1cm of i1](v1);
\coordinate[below left = 1cm of v1, label= left:$\chi$](i2);
\coordinate[above right = 1cm of v1, label=right: $N$] (f1);
\coordinate[below right =  1cm of v1,label=right: $N$] (f2);
\draw[fermionnoarrow] (i1) -- (v1);
\draw[fermionnoarrow] (i2) -- (v1);
\draw[fermionnoarrow] (v1) -- (f1);
\draw[fermionnoarrow] (f2) -- (v1);
\draw[fill=gray] (v1) circle (.3cm);
\end{tikzpicture}
\end{gathered} 
&=&
\begin{gathered}
\begin{tikzpicture}[line width=1.5 pt,node distance=1 cm and 1.5 cm]
\coordinate[label =left: $\chi$] (i1);
\coordinate[below right= 1cm of i1](v1);
\coordinate[ right= 0.5cm of v1,label= above:$J$](vaux);
\coordinate[below left= 1cm of v1, label= left: $\chi$](i2);
\coordinate[right = 1 cm of v1](v2);
\coordinate[above right = 1 cm of v2, label=right: $N$] (f1);
\coordinate[below right =  1 cm of v2,label=right: $N$] (f2);
\draw[fermionnoarrow] (i1) -- (v1);
\draw[fermionnoarrow] (i2) -- (v1);
\draw[scalarnoarrow] (v1) -- (v2);
\draw[fermionnoarrow] (v2) -- (f1);
\draw[fermionnoarrow] (f2) -- (v2);
\draw[fill=black] (v1) circle (.1cm);
\draw[fill=black] (v2) circle (.1cm);
\end{tikzpicture}
\end{gathered} 
+
\begin{gathered}
\begin{tikzpicture}[line width=1.5 pt,node distance=1 cm and 1.5 cm]
\coordinate[label =left: $\chi$] (i1);
\coordinate[below right= 1cm of i1](v1);
\coordinate[ right= 0.5cm of v1,label= above:{$Z_\ell$}](vaux);
\coordinate[below left= 1cm of v1, label= left: $\chi$](i2);
\coordinate[right = 1 cm of v1](v2);
\coordinate[above right = 1 cm of v2, label=right: $N$] (f1);
\coordinate[below right =  1 cm of v2,label=right: $N$] (f2);
\draw[fermionnoarrow] (i1) -- (v1);
\draw[fermionnoarrow] (i2) -- (v1);
\draw[vector] (v1) -- (v2);
\draw[fermionnoarrow] (v2) -- (f1);
\draw[fermionnoarrow] (f2) -- (v2);
\draw[fill=cyan] (v1) circle (.1cm);
\draw[fill=cyan] (v2) circle (.1cm);
\end{tikzpicture}
\end{gathered}
+
\begin{gathered}
\begin{tikzpicture}[line width=1.5 pt,node distance=1 cm and 1.5 cm]
\coordinate[label =left: $\chi$] (i1);
\coordinate[below right= 1cm of i1](v1);
\coordinate[ right= 0.5cm of v1,label= above:$H_i$](vaux);
\coordinate[below left= 1cm of v1, label= left: $\chi$](i2);
\coordinate[right = 1 cm of v1](v2);
\coordinate[above right = 1 cm of v2, label=right: $N$] (f1);
\coordinate[below right =  1 cm of v2,label=right: $N$] (f2);
\draw[fermionnoarrow] (i1) -- (v1);
\draw[fermionnoarrow] (i2) -- (v1);
\draw[scalarnoarrow] (v1) -- (v2);
\draw[fermionnoarrow] (v2) -- (f1);
\draw[fermionnoarrow] (f2) -- (v2);
\draw[fill=black] (v1) circle (.1cm);
\draw[fill=black] (v2) circle (.1cm);
\end{tikzpicture}
\end{gathered}\\
\begin{gathered}
\begin{tikzpicture}[line width=1.5 pt,node distance=1 cm and 1.5 cm]
\coordinate[label = left: $\chi$] (i1);
\coordinate[below right = 1cm of i1](v1);
\coordinate[below left = 1cm of v1, label= left:$\chi$](i2);
\coordinate[above right = 1cm of v1, label=right: $Z_\ell$] (f1);
\coordinate[below right =  1cm of v1,label=right: $Z_\ell$] (f2);
\draw[fermionnoarrow] (i1) -- (v1);
\draw[fermionnoarrow] (i2) -- (v1);
\draw[vector] (v1) -- (f1);
\draw[vector] (f2) -- (v1);
\draw[fill=gray] (v1) circle (.3cm);
\end{tikzpicture}
\end{gathered} 
&=&
\begin{gathered}
\begin{tikzpicture}[line width=1.5 pt,node distance=1 cm and 1.5 cm]
\coordinate[label =left: $\chi$] (i1);
\coordinate[right= 1cm of i1](v1);
\coordinate[below= 0.5cm of v1](vaux);
\coordinate[right = 1cm of v1, label= right:$Z_\ell$](f1);
\coordinate[below = 1 cm of v1](v2);
\coordinate[left = 1 cm of v2, label=left: $\chi$] (i2);
\coordinate[right =  1 cm of v2,label=right: $Z_\ell$] (f2);
\draw[fermionnoarrow] (i1) -- (v1);
\draw[fermionnoarrow] (v1) -- (v2);
\draw[fermionnoarrow] (i2) -- (v2);
\draw[vector] (f2) -- (v2);
\draw[vector] (f1) -- (v1);
\draw[fill=cyan] (v1) circle (.1cm);
\draw[fill=cyan] (v2) circle (.1cm);
\end{tikzpicture}
\end{gathered}
\quad   \! +
\begin{gathered}
\begin{tikzpicture}[line width=1.5 pt,node distance=1 cm and 1.5 cm]
\coordinate[label =left: $\chi$] (i1);
\coordinate[right= 1cm of i1](v1);
\coordinate[below= 0.5cm of v1,label](vaux);
\coordinate[right = 1cm of v1, label= right:$Z_\ell$](f1);
\coordinate[below = 1 cm of v1](v2);
\coordinate[left = 1 cm of v2, label=left: $\chi$] (i2);
\coordinate[right =  1 cm of v2,label=right: $Z_\ell$] (f2);
\draw[fermionnoarrow] (i1) -- (v1);
\draw[fermionnoarrow] (v1) -- (v2);
\draw[fermionnoarrow] (i2) -- (v2);
\draw[vector] (f2) -- (v1);
\draw[vector] (f1) -- (v2);
\draw[fill=cyan] (v1) circle (.1cm);
\draw[fill=cyan] (v2) circle (.1cm);
\end{tikzpicture}
\end{gathered}
+
\begin{gathered}
\begin{tikzpicture}[line width=1.5 pt,node distance=1 cm and 1.5 cm]
\coordinate[label =left: $\chi$] (i1);
\coordinate[below right= 1cm of i1](v1);
\coordinate[ right= 0.5cm of v1,label= above:{$H_i$}](vaux);
\coordinate[below left= 1cm of v1, label= left: $\chi$](i2);
\coordinate[right = 1 cm of v1](v2);
\coordinate[above right = 1 cm of v2, label=right: $Z_\ell$] (f1);
\coordinate[below right =  1 cm of v2,label=right: $Z_\ell$] (f2);
\draw[fermionnoarrow] (i1) -- (v1);
\draw[fermionnoarrow] (i2) -- (v1);
\draw[scalarnoarrow] (v1) -- (v2);
\draw[vector] (v2) -- (f1);
\draw[vector] (f2) -- (v2);
\draw[fill=black] (v1) circle (.1cm);
\draw[fill=cyan] (v2) circle (.1cm);
\end{tikzpicture}
\end{gathered}
\\
\begin{gathered}
\begin{tikzpicture}[line width=1.5 pt,node distance=1 cm and 1.5 cm]
\coordinate[label = left: $\chi$] (i1);
\coordinate[below right = 1cm of i1](v1);
\coordinate[below left = 1cm of v1, label= left:$\chi$](i2);
\coordinate[above right = 1cm of v1, label=right: $Z_\ell$] (f1);
\coordinate[below right =  1cm of v1,label=right: $H_i$] (f2);
\draw[fermionnoarrow] (i1) -- (v1);
\draw[fermionnoarrow] (i2) -- (v1);
\draw[vector] (v1) -- (f1);
\draw[scalarnoarrow] (f2) -- (v1);
\draw[fill=gray] (v1) circle (.3cm);
\end{tikzpicture}
\end{gathered} 
&=&
\begin{gathered}
\begin{tikzpicture}[line width=1.5 pt,node distance=1 cm and 1.5 cm]
\coordinate[label =left: $\chi$] (i1);
\coordinate[right= 1cm of i1](v1);
\coordinate[below= 0.5cm of v1](vaux);
\coordinate[right = 1cm of v1, label= right:$Z_\ell$](f1);
\coordinate[below = 1 cm of v1](v2);
\coordinate[left = 1 cm of v2, label=left: $\chi$] (i2);
\coordinate[right =  1 cm of v2,label=right: $H_i$] (f2);
\draw[fermionnoarrow] (i1) -- (v1);
\draw[fermionnoarrow] (v1) -- (v2);
\draw[fermionnoarrow] (i2) -- (v2);
\draw[scalarnoarrow] (f2) -- (v2);
\draw[vector] (f1) -- (v1);
\draw[fill=cyan] (v1) circle (.1cm);
\draw[fill=black] (v2) circle (.1cm);
\end{tikzpicture}
\end{gathered}
\quad   \! +
\begin{gathered}
\begin{tikzpicture}[line width=1.5 pt,node distance=1 cm and 1.5 cm]
\coordinate[label =left: $\chi$] (i1);
\coordinate[right= 1cm of i1](v1);
\coordinate[below= 0.5cm of v1,label](vaux);
\coordinate[right = 1cm of v1, label= right:$Z_\ell$](f1);
\coordinate[below = 1 cm of v1](v2);
\coordinate[left = 1 cm of v2, label=left: $\chi$] (i2);
\coordinate[right =  1 cm of v2,label=right: $H_i$] (f2);
\draw[fermionnoarrow] (i1) -- (v1);
\draw[fermionnoarrow] (v1) -- (v2);
\draw[fermionnoarrow] (i2) -- (v2);
\draw[scalarnoarrow] (f2) -- (v1);
\draw[vector] (f1) -- (v2);
\draw[fill=black] (v1) circle (.1cm);
\draw[fill=cyan] (v2) circle (.1cm);
\end{tikzpicture}
\end{gathered}
+
\begin{gathered}
\begin{tikzpicture}[line width=1.5 pt,node distance=1 cm and 1.5 cm]
\coordinate[label =left: $\chi$] (i1);
\coordinate[below right= 1cm of i1](v1);
\coordinate[ right= 0.5cm of v1,label= above:{$Z_\ell$}](vaux);
\coordinate[below left= 1cm of v1, label= left: $\chi$](i2);
\coordinate[right = 1 cm of v1](v2);
\coordinate[above right = 1 cm of v2, label=right: $Z_\ell$] (f1);
\coordinate[below right =  1 cm of v2,label=right: $H_i$] (f2);
\draw[fermionnoarrow] (i1) -- (v1);
\draw[fermionnoarrow] (i2) -- (v1);
\draw[vector] (v1) -- (v2);
\draw[vector] (v2) -- (f1);
\draw[scalarnoarrow] (f2) -- (v2);
\draw[fill=cyan] (v1) circle (.1cm);
\draw[fill=cyan] (v2) circle (.1cm);
\end{tikzpicture}
\end{gathered}
\\
\begin{gathered}
\begin{tikzpicture}[line width=1.5 pt,node distance=1 cm and 1.5 cm]
\coordinate[label = left: $\chi$] (i1);
\coordinate[below right = 1cm of i1](v1);
\coordinate[below left = 1cm of v1, label= left:$\chi$](i2);
\coordinate[above right = 1cm of v1, label=right: $J$] (f1);
\coordinate[below right =  1cm of v1,label=right: $J$] (f2);
\draw[fermionnoarrow] (i1) -- (v1);
\draw[fermionnoarrow] (i2) -- (v1);
\draw[scalarnoarrow] (v1) -- (f1);
\draw[scalarnoarrow] (f2) -- (v1);
\draw[fill=gray] (v1) circle (.3cm);
\end{tikzpicture}
\end{gathered} 
&=& 
\begin{gathered}
\begin{tikzpicture}[line width=1.5 pt,node distance=1 cm and 1.5 cm]
\coordinate[label =left: $\chi$] (i1);
\coordinate[right= 1cm of i1](v1);
\coordinate[below= 0.5cm of v1](vaux);
\coordinate[right = 1cm of v1, label= right:$J$](f1);
\coordinate[below = 1 cm of v1](v2);
\coordinate[left = 1 cm of v2, label=left: $\chi$] (i2);
\coordinate[right =  1 cm of v2,label=right: $J$] (f2);
\draw[fermionnoarrow] (i1) -- (v1);
\draw[fermionnoarrow] (v1) -- (v2);
\draw[fermionnoarrow] (i2) -- (v2);
\draw[scalarnoarrow] (f2) -- (v2);
\draw[scalarnoarrow] (f1) -- (v1);
\draw[fill=black] (v1) circle (.1cm);
\draw[fill=black] (v2) circle (.1cm);
\end{tikzpicture}
\end{gathered}
\quad   \! +
\begin{gathered}
\begin{tikzpicture}[line width=1.5 pt,node distance=1 cm and 1.5 cm]
\coordinate[label =left: $\chi$] (i1);
\coordinate[right= 1cm of i1](v1);
\coordinate[below= 0.5cm of v1,label](vaux);
\coordinate[right = 1cm of v1, label= right:$J$](f1);
\coordinate[below = 1 cm of v1](v2);
\coordinate[left = 1 cm of v2, label=left: $\chi$] (i2);
\coordinate[right =  1 cm of v2,label=right: $J$] (f2);
\draw[fermionnoarrow] (i1) -- (v1);
\draw[fermionnoarrow] (v1) -- (v2);
\draw[fermionnoarrow] (i2) -- (v2);
\draw[scalarnoarrow] (f2) -- (v1);
\draw[scalarnoarrow] (f1) -- (v2);
\draw[fill=black] (v1) circle (.1cm);
\draw[fill=black] (v2) circle (.1cm);
\end{tikzpicture}
\end{gathered}
\\
\begin{gathered}
\begin{tikzpicture}[line width=1.5 pt,node distance=1 cm and 1.5 cm]
\coordinate[label = left: $\chi$] (i1);
\coordinate[below right = 1cm of i1](v1);
\coordinate[below left = 1cm of v1, label= left:$\chi$](i2);
\coordinate[above right = 1cm of v1, label=right: $H_i$] (f1);
\coordinate[below right =  1cm of v1,label=right: $H_j$] (f2);
\draw[fermionnoarrow] (i1) -- (v1);
\draw[fermionnoarrow] (i2) -- (v1);
\draw[scalarnoarrow] (v1) -- (f1);
\draw[scalarnoarrow] (f2) -- (v1);
\draw[fill=gray] (v1) circle (.3cm);
\end{tikzpicture}
\end{gathered} 
&=&
\begin{gathered}
\begin{tikzpicture}[line width=1.5 pt,node distance=1 cm and 1.5 cm]
\coordinate[label =left: $\chi$] (i1);
\coordinate[right= 1cm of i1](v1);
\coordinate[below= 0.5cm of v1](vaux);
\coordinate[right = 1cm of v1, label= right:$H_i$](f1);
\coordinate[below = 1 cm of v1](v2);
\coordinate[left = 1 cm of v2, label=left: $\chi$] (i2);
\coordinate[right =  1 cm of v2,label=right: $H_j$] (f2);
\draw[fermionnoarrow] (i1) -- (v1);
\draw[fermionnoarrow] (v1) -- (v2);
\draw[fermionnoarrow] (i2) -- (v2);
\draw[scalarnoarrow] (f2) -- (v2);
\draw[scalarnoarrow] (f1) -- (v1);
\draw[fill=black] (v1) circle (.1cm);
\draw[fill=black] (v2) circle (.1cm);
\end{tikzpicture}
\end{gathered}
\quad   \! +
\begin{gathered}
\begin{tikzpicture}[line width=1.5 pt,node distance=1 cm and 1.5 cm]
\coordinate[label =left: $\chi$] (i1);
\coordinate[right= 1cm of i1](v1);
\coordinate[below= 0.5cm of v1,label](vaux);
\coordinate[right = 1cm of v1, label= right:$H_i$](f1);
\coordinate[below = 1 cm of v1](v2);
\coordinate[left = 1 cm of v2, label=left: $\chi$] (i2);
\coordinate[right =  1 cm of v2,label=right: $H_j$] (f2);
\draw[fermionnoarrow] (i1) -- (v1);
\draw[fermionnoarrow] (v1) -- (v2);
\draw[fermionnoarrow] (i2) -- (v2);
\draw[scalarnoarrow] (f2) -- (v1);
\draw[scalarnoarrow] (f1) -- (v2);
\draw[fill=black] (v1) circle (.1cm);
\draw[fill=black] (v2) circle (.1cm);
\end{tikzpicture}
\end{gathered}
+
\begin{gathered}
\begin{tikzpicture}[line width=1.5 pt,node distance=1 cm and 1.5 cm]
\coordinate[label =left: $\chi$] (i1);
\coordinate[below right= 1cm of i1](v1);
\coordinate[ right= 0.5cm of v1,label= above:{$h$,$H_k$}](vaux);
\coordinate[below left= 1cm of v1, label= left: $\chi$](i2);
\coordinate[right = 1 cm of v1](v2);
\coordinate[above right = 1 cm of v2, label=right: $H_i$] (f1);
\coordinate[below right =  1 cm of v2,label=right: $H_j$] (f2);
\draw[fermionnoarrow] (i1) -- (v1);
\draw[fermionnoarrow] (i2) -- (v1);
\draw[scalarnoarrow] (v1) -- (v2);
\draw[scalarnoarrow] (v2) -- (f1);
\draw[scalarnoarrow] (f2) -- (v2);
\draw[fill=black] (v1) circle (.1cm);
\draw[fill=black] (v2) circle (.1cm);
\end{tikzpicture}
\end{gathered}\\
\begin{gathered}
\begin{tikzpicture}[line width=1.5 pt,node distance=1 cm and 1.5 cm]
\coordinate[label = left: $\chi$] (i1);
\coordinate[below right = 1cm of i1](v1);
\coordinate[below left = 1cm of v1, label= left:$\chi$](i2);
\coordinate[above right = 1cm of v1, label=right: $Z_\ell$] (f1);
\coordinate[below right =  1cm of v1,label=right: $J$] (f2);
\draw[fermionnoarrow] (i1) -- (v1);
\draw[fermionnoarrow] (i2) -- (v1);
\draw[vector] (v1) -- (f1);
\draw[scalarnoarrow] (f2) -- (v1);
\draw[fill=gray] (v1) circle (.3cm);
\end{tikzpicture}
\end{gathered} 
&=&
\begin{gathered}
\begin{tikzpicture}[line width=1.5 pt,node distance=1 cm and 1.5 cm]
\coordinate[label =left: $\chi$] (i1);
\coordinate[right= 1cm of i1](v1);
\coordinate[below= 0.5cm of v1](vaux);
\coordinate[right = 1cm of v1, label= right:$Z_\ell$](f1);
\coordinate[below = 1 cm of v1](v2);
\coordinate[left = 1 cm of v2, label=left: $\chi$] (i2);
\coordinate[right =  1 cm of v2,label=right: $J$] (f2);
\draw[fermionnoarrow] (i1) -- (v1);
\draw[fermionnoarrow] (v1) -- (v2);
\draw[fermionnoarrow] (i2) -- (v2);
\draw[scalarnoarrow] (f2) -- (v2);
\draw[vector] (f1) -- (v1);
\draw[fill=cyan] (v1) circle (.1cm);
\draw[fill=black] (v2) circle (.1cm);
\end{tikzpicture}
\end{gathered}
\quad   \! +
\begin{gathered}
\begin{tikzpicture}[line width=1.5 pt,node distance=1 cm and 1.5 cm]
\coordinate[label =left: $\chi$] (i1);
\coordinate[right= 1cm of i1](v1);
\coordinate[below= 0.5cm of v1,label](vaux);
\coordinate[right = 1cm of v1, label= right:$Z_\ell$](f1);
\coordinate[below = 1 cm of v1](v2);
\coordinate[left = 1 cm of v2, label=left: $\chi$] (i2);
\coordinate[right =  1 cm of v2,label=right: $J$] (f2);
\draw[fermionnoarrow] (i1) -- (v1);
\draw[fermionnoarrow] (v1) -- (v2);
\draw[fermionnoarrow] (i2) -- (v2);
\draw[scalarnoarrow] (f2) -- (v1);
\draw[vector] (f1) -- (v2);
\draw[fill=black] (v1) circle (.1cm);
\draw[fill=cyan] (v2) circle (.1cm);
\end{tikzpicture}
\end{gathered}
+
\begin{gathered}
\begin{tikzpicture}[line width=1.5 pt,node distance=1 cm and 1.5 cm]
\coordinate[label =left: $\chi$] (i1);
\coordinate[below right= 1cm of i1](v1);
\coordinate[ right= 0.5cm of v1,label= above:$H_i$](vaux);
\coordinate[below left= 1cm of v1, label= left: $\chi$](i2);
\coordinate[right = 1 cm of v1](v2);
\coordinate[above right = 1 cm of v2, label=right: $Z_\ell$] (f1);
\coordinate[below right =  1 cm of v2,label=right: $J$] (f2);
\draw[fermionnoarrow] (i1) -- (v1);
\draw[fermionnoarrow] (i2) -- (v1);
\draw[scalarnoarrow] (v1) -- (v2);
\draw[vector] (v2) -- (f1);
\draw[scalarnoarrow] (f2) -- (v2);
\draw[fill=black] (v1) circle (.1cm);
\draw[fill=cyan] (v2) circle (.1cm);
\end{tikzpicture}
\end{gathered}\\
\begin{gathered}
\begin{tikzpicture}[line width=1.5 pt,node distance=1 cm and 1.5 cm]
\coordinate[label = left: $\chi$] (i1);
\coordinate[below right = 1cm of i1](v1);
\coordinate[below left = 1cm of v1, label= left:$\chi$](i2);
\coordinate[above right = 1cm of v1, label=right: $H_i$] (f1);
\coordinate[below right =  1cm of v1,label=right: $J$] (f2);
\draw[fermionnoarrow] (i1) -- (v1);
\draw[fermionnoarrow] (i2) -- (v1);
\draw[scalarnoarrow] (v1) -- (f1);
\draw[scalarnoarrow] (f2) -- (v1);
\draw[fill=gray] (v1) circle (.3cm);
\end{tikzpicture}
\end{gathered} 
&=&
\begin{gathered}
\begin{tikzpicture}[line width=1.5 pt,node distance=1 cm and 1.5 cm]
\coordinate[label =left: $\chi$] (i1);
\coordinate[right= 1cm of i1](v1);
\coordinate[below= 0.5cm of v1](vaux);
\coordinate[right = 1cm of v1, label= right:$H_i$](f1);
\coordinate[below = 1 cm of v1](v2);
\coordinate[left = 1 cm of v2, label=left: $\chi$] (i2);
\coordinate[right =  1 cm of v2,label=right: $J$] (f2);
\draw[fermionnoarrow] (i1) -- (v1);
\draw[fermionnoarrow] (v1) -- (v2);
\draw[fermionnoarrow] (i2) -- (v2);
\draw[scalarnoarrow] (f2) -- (v2);
\draw[scalarnoarrow] (f1) -- (v1);
\draw[fill=black] (v1) circle (.1cm);
\draw[fill=black] (v2) circle (.1cm);
\end{tikzpicture}
\end{gathered}
\quad   \! +
\begin{gathered}
\begin{tikzpicture}[line width=1.5 pt,node distance=1 cm and 1.5 cm]
\coordinate[label =left: $\chi$] (i1);
\coordinate[right= 1cm of i1](v1);
\coordinate[below= 0.5cm of v1,label](vaux);
\coordinate[right = 1cm of v1, label= right:$H_i$](f1);
\coordinate[below = 1 cm of v1](v2);
\coordinate[left = 1 cm of v2, label=left: $\chi$] (i2);
\coordinate[right =  1 cm of v2,label=right: $J$] (f2);
\draw[fermionnoarrow] (i1) -- (v1);
\draw[fermionnoarrow] (v1) -- (v2);
\draw[fermionnoarrow] (i2) -- (v2);
\draw[scalarnoarrow] (f2) -- (v1);
\draw[scalarnoarrow] (f1) -- (v2);
\draw[fill=black] (v1) circle (.1cm);
\draw[fill=black] (v2) circle (.1cm);
\end{tikzpicture}
\end{gathered}
+
\begin{gathered}
\begin{tikzpicture}[line width=1.5 pt,node distance=1 cm and 1.5 cm]
\coordinate[label =left: $\chi$] (i1);
\coordinate[below right= 1cm of i1](v1);
\coordinate[ right= 0.5cm of v1,label= above:$Z_\ell$](vaux);
\coordinate[below left= 1cm of v1, label= left: $\chi$](i2);
\coordinate[right = 1 cm of v1](v2);
\coordinate[above right = 1 cm of v2, label=right: $H_i$] (f1);
\coordinate[below right =  1 cm of v2,label=right: $J$] (f2);
\draw[fermionnoarrow] (i1) -- (v1);
\draw[fermionnoarrow] (i2) -- (v1);
\draw[vector] (v1) -- (v2);
\draw[scalarnoarrow] (v2) -- (f1);
\draw[scalarnoarrow] (f2) -- (v2);
\draw[fill=cyan] (v1) circle (.1cm);
\draw[fill=cyan] (v2) circle (.1cm);
\end{tikzpicture}
\end{gathered}
\label{graphs}
\end{eqnarray*}
\end{figure}

\newpage
\section{SOMMERFELD ENHANCEMENT}
\label{SecSom}
When the new Higgses in the theory are light and one has large Yukawa couplings between the dark matter and the new Higgses, one can expect a large Sommerfeld enhancement~\cite{Hisano_2005}. To determine the enhancement factor for a Yukuwa potential, one needs to solve the Schr\"odinger equation numerically. The enhancement factor can also be obtained by approximating the Yukawa potential to a Hulthen potential and the resulting analytic approximation is given by~\cite{Slatyer_2010,Cassel_2010}
\begin{equation}
     S = \frac{\pi}{\epsilon_{\nu}}\frac{\sinh{\left(\frac{2 \pi \epsilon_\nu}{\pi^2 \epsilon_\phi/6}\right)}}{\cosh{\left(\frac{2 \pi \epsilon_\nu}{\pi^2 \epsilon_\phi/6}\right)} - \cos{\left(2 \pi \sqrt{\frac{1}{\pi^2 \epsilon_\phi/6}-\frac{\epsilon_\nu^2}{(\pi^2 \epsilon_\phi/6)^2}} \right)}}
 \end{equation}
 In the minimal theory studied in the previous section the SM neutrinos are Dirac fermions. In this case, we have
 \begin{equation}
 \epsilon_\nu = v/\alpha_\chi, \ \epsilon_\phi = M_{h_\ell}/(\alpha_\chi M_{\chi}), \ \alpha_\chi = \frac{1}{4 \pi}\left(\frac{3g_\ell M_\chi}{2 M_{Z_\ell}}\right)^2,
 \end{equation}
 where $v$ is the dark matter velocity in the center of mass frame and $M_{h_\ell}$ is the mediator (Higgs) mass. We have discussed above the implementation of the seesaw mechanism for Majorana neutrino masses. In this scenario, we two mediators, $H_1$, and $H_2$, contributing to the Sommerfeld enhancement. In this case, the Hulthen potential can be written as
 $$V(r)=\frac{\alpha_{H_1}\delta_{H_1}e^{-\delta_{H_1}r}}{1-e^{-\delta_{H_1 }r}} +\frac{\alpha_{H_2}\delta_{H_2}e^{-\delta_{H_2}r}}{1-e^{-\delta_{H_2 }r}}.$$
 Here $\delta_{H_i} = \pi^2 M_{H_i}/6$.
 In the limit, $M_{H_1}\approx M_{H_2}$, 
 the Hulthen potential becomes
 $$V(r)=\frac{\alpha_{\chi_M} \ \delta \ e^{-\delta r}}{1-e^{-\delta r}},$$
 where $\alpha_{\chi_M}=\alpha_{H_1}+\alpha_{H_2}$, $\delta=\delta_{H_1}=\delta_{H_2}$ and $\alpha_{H_i}= \frac{1}{4\pi}\left(\frac{3 M_\chi g_\ell}{2 M_{Z_\ell}} \frac{U_{2i}}{\sin{\beta}}\right)^2$. Thus, the enhancement factor follows Eq.\textcolor{blue}{C1}.
 This non-perturbative effect might play a role during freezing out, for details see Refs.~\cite{Cirelli_2007,Feng_2010}. Notice that in our studies the Yukawa coupling is proportional to the dark matter mass. Therefore, one can expect a large enhancement factor for a heavy dark matter. 
\section{CROSS SECTIONS}
The cross section for $\bar{\nu}_R \nu_R \to Z_\ell^* \to \bar{f} f$ is given by
\begin{eqnarray}
    \sigma_{\bar{\nu}_R \nu_R \to \bar{f} f} (s)= \frac{g_{\ell}^4}{12 \pi \sqrt{s}} \frac{1}{(s-M_{Z_{\ell}}^2)^2+ \Gamma_{Z_{\ell}}^2 M_{Z_{\ell}}^2} \sqrt{s-M_f^2}\hspace{0.1 cm}(s+2M_f^2) \sum_f N_f^C n_f^2.
    \label{nuRnuRff}
\end{eqnarray}
The dark matter annihilation cross section into two photons is given by
\begin{equation}
 \sigma v ({{\chi}\chi \to \gamma \gamma}) =\frac{\alpha^2}{\pi^3} \frac{g_\ell^4 n_\chi^2 M_\chi^2}{M_{Z_\ell}^4}\frac{(4M_\chi^2 - M_{Z_\ell}^2)^2}{(4M_\chi^2-M_{Z_\ell}^2)^2+\Gamma_{Z_\ell}^2M_{Z_\ell}^2} \left | \sum_{f_+} N_c^{F} n_A^{F}Q_{F}^2M_{F}^2 C_0^\gamma \right |^2,
\label{Gamma1}
\end{equation}
where the loop function can be written as 
\begin{eqnarray}
    C_0^\gamma &=&\frac{1}{2s} ln^2 \left( \frac{\sqrt{1 - 4 M_F^2/s}-1}{\sqrt{1 - 4 M_F^2/s}+1} \right).
\end{eqnarray}
This loop funtion is evaluated at $s= 4 M_\chi^2$.

The dark matter annihilation cross section into a photon and $Z$ is given by
\begin{equation}
\sigma v (\chi \chi \to \gamma Z) =\frac{\alpha^2 \, g_\ell^4 n_\chi^2 }{32 \pi^3 \sin^2 2\theta_W}\frac{(4M_\chi^2-M_Z^2)^3}{ (4M_\chi^2-M_{Z_\ell}^2)^2+\Gamma_{Z_\ell}^2 M_{Z_\ell}^2}\frac{(M_{Z_\ell}^2-4M_\chi^2)^2}{M_\chi^4M_{Z_\ell}^4} \left| \sum_f N_c^{F} Q_{F}  n_A^{F} g_{V}^{F} 2M_{F}^2 C_0^Z \right|^2,
\label{chichiZgamma}
\end{equation}
and the loop function is 
\begin{eqnarray}
    C_0^Z &=&\frac{1}{2(M_Z^2 -s)} \left( ln^2 \left( \frac{\sqrt{1 - 4 M_F^2/M_Z^2}-1}{\sqrt{1 - 4 M_F^2/M_Z^2}+1} \right) - ln^2 \left( \frac{\sqrt{1 - 4 M_F^2/s}-1}{\sqrt{1 - 4 M_F^2/s}+1} \right) \right).
\end{eqnarray}
The coefficients $A$ and $B$ in Eq.(\ref{FSR1}) are given by
\begin{eqnarray}
A&=&- 9 \pi \, \alpha \, g_\ell^4    (M_{Z_\ell}^2-4M_\chi^2)^2 \frac{\left(2(E_\ell-M_\chi)(E_\ell + E_\gamma -M_\chi)-3M_\ell^2 \right) }{M_{Z_\ell}^2(E_\ell-M_\chi)^2 ((4M_\chi^2-M_{Z_\ell}^2)^2+\Gamma_{Z_\ell}^2M_{Z_\ell}^2)}, 
\label{A}
\end{eqnarray}
and
\begin{eqnarray}
B&=& 9 \pi \, \alpha \,  g_\ell^4 M_\chi^2 Q_\ell^2  \times \nonumber \\
&&  \frac{ \left(2E_\ell M_\chi (E_\gamma^2-3 E_\gamma M_\chi + 2 M_\chi^2)-2 E_\ell^4 - 2 E_\ell^3 ( E_\gamma - 2 M_\chi) -E_\ell^2(E_\gamma^2-6E_\gamma M_\chi + 6M_\chi^2)-2M_\chi^2(E_\gamma - M_\chi)^2\right)}
{E_\ell^2 (E_\ell - M_\chi) (E_\ell+E_\gamma-M_\chi)((4M_\chi^2-M_{Z_\ell}^2)^2+\Gamma_{Z_\ell}^2M_{Z_\ell}^2)}.
\nonumber \\
\label{B}
\end{eqnarray}
\end{widetext}
\bibliography{refs}

\end{document}